\documentclass[useAMS,usenatbib,a4paper]{mn2e}

\usepackage[left=2cm,top=2cm,right=2cm,bottom=2cm]{geometry}
\usepackage{amssymb,epsfig,color}
\usepackage{amsmath}
\usepackage{graphicx}
\usepackage[caption=false]{caption}
\usepackage[font=footnotesize]{subfig}
\usepackage{multirow}
\usepackage{xspace}
\usepackage{hyperref}
\usepackage{rotating}

\newcommand{\lan}{\langle}
\newcommand{\ran}{\rangle}

\newcommand{\avg}[1]{\left\langle #1 \right\rangle}
\newcommand{\dg}{^{\circ}}

\newcommand{\fign}[1]{Figure~\ref{#1}\xspace}

\title[The C-Band All-Sky Survey (C-BASS): Northern Receiver]{The C-Band All-Sky Survey (C-BASS):
Design and implementation of the northern receiver}
\author[O.$\,$G. King et al.]{O.$\,$G. King$^{1,2}$\thanks{E-mail: ogk@astro.caltech.edu}, Michael$\,$E.
Jones$^{2}$, E.$\,$J. Blackhurst$^{3}$, C. Copley$^{2}$, R.$\,$J. Davis$^{3}$, 
\newauthor C. Dickinson$^{3}$, C.$\,$M. Holler$^{4,2}$, M.$\,$O. Irfan$^{3}$, J.$\,$J. John$^{2}$,
J.$\,$P. Leahy$^{3}$, J. Leech$^{2}$,
\newauthor S.$\,$J.$\,$C. Muchovej$^{1}$, T.$\,$J. Pearson$^{1}$, M.$\,$A. Stevenson$^{1}$, and
Angela$\,$C. Taylor$^{2}$ \\
$^{1}$California Institute of Technology, Pasadena CA 91125, USA\\
$^{2}$Sub-department of Astrophysics, University of Oxford, Denys Wilkinson Building, Keble Road,
Oxford, OX1 3RH, UK\\
$^{3}$Jodrell Bank Centre for Astrophysics, School of Physics \& Astronomy, The University of Manchester, \\Oxford Road, Manchester, M13 9PL, UK \\
$^{4}$Hochschule Esslingen, Kanalstra{\ss}e 33, Esslingen 73728, Germany}

\begin{document}

\date{Accepted XXX. Received YYY; in original form ZZZ}

\pagerange{\pageref{firstpage}--\pageref{lastpage}} \pubyear{2013}

\maketitle

\label{firstpage}

\begin{abstract}
The C-Band All-Sky Survey (C-BASS) is a project to map the full sky in
total intensity and linear polarization at 5\,GHz. The northern
component of the survey uses a broadband single-frequency analogue
receiver fitted to a 6.1-m telescope at the Owens Valley Radio
Observatory in California, USA.  The receiver architecture combines a
continuous-comparison radiometer and a correlation polarimeter in a
single receiver for stable simultaneous measurement of both total
intensity and linear polarization, using custom-designed analogue
receiver components. The continuous-comparison radiometer measures the
temperature difference between the sky and temperature-stabilized cold
electrical reference loads. A cryogenic front-end is used to minimize
receiver noise, with a system temperature of $\approx 30\,$K in both linear
polarization and total intensity.  Custom cryogenic notch filters are
used to counteract man-made radio frequency interference. The
radiometer $1/f$ noise is dominated by atmospheric fluctuations, while
the polarimeter achieves a $1/f$ noise knee frequency of 10\,mHz, similar to
the telescope azimuthal scan frequency.

\end{abstract}

\begin{keywords}
instrumentation: polarimeters -- Galaxy: general -- radio continuum: general.
\end{keywords}

\section{Introduction}

The C-Band All-Sky Survey (C-BASS)\footnote{\url{http://www.astro.caltech.edu/cbass/}} is an 
experiment to map the entire sky in total
intensity and linear polarization at $5\,$GHz. C-BASS aims to provide high signal-to-noise
ratio maps of the polarization of Galactic synchrotron radiation,
largely uncorrupted by Faraday rotation. The survey is being conducted
in two parts: for the northern sky, a new instrument has been deployed
on a 6.1-m antenna at the Owens Valley Radio Observatory (OVRO),
California, USA, while for the southern sky a similar instrument is used on
a 7.6-m telescope at the MeerKAT \citep{2012AfrSk..16..101B} support site near Carnarvon, South
Africa. The two systems are matched to each give the same resolution
($0.\!\dg73$, \citealt{Holler:2012dp}) and similar system temperatures of $\approx 30\,$K, and the
data will be merged to form a single set of all-sky images.

The primary purpose of the C-BASS experiment is to assist in the measurement
of the polarized cosmic microwave background radiation
(CMB) \citep{King:2010gs}. Measuring the $B$-mode polarization signal requires separation
of the CMB from foreground emission, which at 5\,GHz is dominated by diffuse
Galactic synchrotron emission, to greater accuracy than is
possible with our current understanding of Galactic
foregrounds \citep{2007ApJ...665..355K,2011ApJS..192...15G,2011MNRAS.418..888M}. C-BASS data are expected to improve the
accuracy of foreground subtraction in, for instance, template fitting
analyses \citep{2012MNRAS.422.3617G}. 

The sensitivity requirement for the survey is set by requiring that
there be a 5-sigma detection of polarization in at least 90\% of the
sky.  Interpolating between existing maps at 1.41\,GHz
\citep{2006A&A...448..411W} and 23\,GHz \citep{Page:2007ce} we
obtained an estimate that 90\% of the sky pixels would have a polarized intensity of 0.5\,mK or
greater, giving a sensitivity requirement of $<0.1\,$mK/beam in linear
polarization. The instrument design will reach the same thermal noise
level in intensity, although we expect the intensity maps to be
confusion limited at a higher level, $\approx 1\,$mK.  

Achieving this sensitivity is a balance between bandwidth, system temperature, and integration
time as dictated by the radiometer equation \citep{Kraus:1986}:
\begin{align}
 \sigma = \frac{T_{\rm sys}}{\sqrt{\Delta \nu \tau}},
\end{align}
where $\sigma$ is the noise level, $T_{\rm sys}$ is the system temperature, $\Delta\nu$ is the
bandwidth, and $\tau$ is the integration time. While a broader bandwidth would make the
target sensitivity easier to achieve, interference from man-made radio transmissions limited our
effective bandwidth to $\approx 500\,$MHz in the northern survey, and a consequent thermal noise level of
1.3\,mK\,s$^{-1/2}$ given a 30\,K system temperature. Reaching this noise level requires
a low level of $1/f$ noise in the receiver: in a conventional radiometer the $1/f$ noise would
exceed the thermal noise level set by the radiometer equation over the time-scale of the
measurement. We overcome the $1/f$ noise problem by designing a receiver that suppresses $1/f$
noise, and using a destriping map-making algorithm \citep{Sutton2009} to remove the $1/f$ noise
that remains.

The C-BASS receiver is designed to measure both total intensity and
linear polarization simultaneously, with minimal systematic error, in a single frequency channel
between 4.5 and $5.5\,$GHz. To achieve this we have developed a hybrid
receiver design in which the total intensity is measured in a
continuous-comparison fashion by comparing the sky temperature against
a cold electrical reference load, similar to the \emph{Planck} LFI receivers \citep{PlanckLFI}.
The linear polarization is measured by correlating orthogonal circular polarization signals from
the horn. This architecture suppresses the instabilities due to $1/f$
noise in the receiver in both the intensity and polarization
measurements. The C-BASS receiver is unique in its combination of a 
continuous-comparison radiometer and a correlation polarimeter in a
single instrument. 
Some polarimeters, e.g.
QUIET \citep{2010SPIE.7741E..38B,2012arXiv1207.5562Q} and GEM
\citep{2011ExA....30...23B}, measure polarization the same way that we do,
but do not measure total intensity with the $1/f$ noise suppression of the continuous comparison
architecture\footnote{The QUIET receiver had some pixels connected in
  a differential-temperature assembly that measured the temperature
  difference between two horns. However, this introduces spatial
  filtering, as discussed in Section~\ref{sec:radiometer_design}.}. Similarly, some continuous-comparison receivers, e.g. {\em Planck} LFI
\citep{PlanckLFI}, are able to measure one component of the linear polarization vector by
differencing the power in orthogonal linear modes, but are only able to measure the full linear
Stokes vector by rotating or changing the orientation of the receiver with respect to the sky. The hybrid
architecture used in the C-BASS receiver does have several disadvantages: four receiver gain chains
are needed instead of two, hence greater cost, and the addition of a $180^{\circ}$ hybrid to the
signal path before the first low-noise amplifier degrades the sensitivity of the instrument.

In this paper we describe the analogue receiver built for the northern
survey. 
The northern and southern receivers have the same architecture and share identical cryogenic
front-ends, but the southern receiver was built later and took advantage of advances in digital
processing hardware to replace the rest of the receiver with real-time digital processing (C.
Copley et al., in prep.).
The project as a whole is described in the project 
paper (M.E. Jones et al., in prep.) and the on-sky commissioning of the northern instrument
is described in S.J.C. Muchovej et al., in prep.
In Section~\ref{sec:instrument_design} we describe the overall
architecture of the system. Sections~\ref{sec:implementation_cryostat} and
\ref{sec:implementation_polarimeter} describe in detail the
implementations of the cryogenic receiver and warm electronics
respectively. We describe the performance of individual sections of
the receiver throughout the text, and in Section~\ref{sec:instrument_performance} present
performance results for the full receiver. We begin each section with a qualitative outline,
followed by a detailed description.

\section{Instrument Design} \label{sec:instrument_design}

\begin{figure*}
 \centering
 \includegraphics[angle=90,width=15cm]{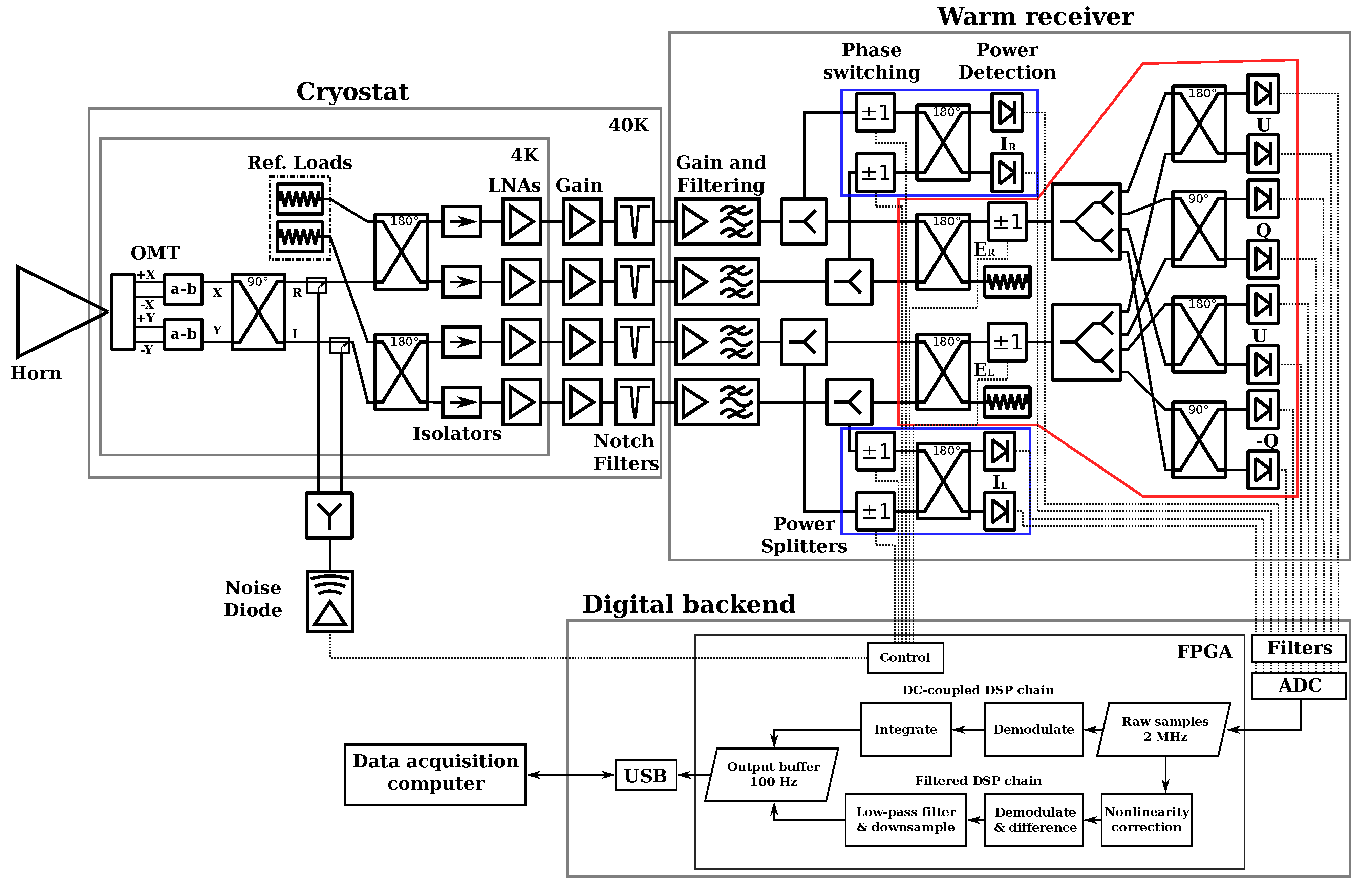}
\caption{A simplified schematic diagram of the C-BASS receiver. Left
  and right circular polarizations (L and R respectively) are
  extracted from the feed-horn and OMT and combined with
  temperature-controlled reference loads using $180\dg$ hybrids. After
  gain and filtering a bank of $180\dg$ hybrids correlates the total
  intensity signals (blue boxes) and recovers the circular
  polarization voltages, which are fed in to the correlation
  polarimeter section (red box). The detected powers are digitized
  and processed by digital signal processing (DSP) chains in an
  FPGA-based digital readout system and then stored on disk.}
 \label{fig:cbass_receiver_diagram}
\end{figure*}

The northern C-BASS instrument is a combination of a
continuous-comparison radiometer and a correlation polarimeter. A
simplified schematic diagram of the instrument is shown in
Fig.~\ref{fig:cbass_receiver_diagram}. The instrument consists of a
cryogenic receiver followed by a warm radiometer/polarimeter and then
a digital backend. In the cryogenic receiver, orthogonal linear
polarizations are extracted from the sky signal by the orthomode
transducer (OMT). These pass through a
$90^{\circ}$ hybrid (included in the OMT block in
Fig.~\ref{fig:cbass_receiver_diagram}) to form orthogonal circular
polarizations $E_{L}$ and $E_{R}$. A calibration signal from a noise
diode is weakly coupled in to each circular polarization using $-30$ dB
directional couplers. The circular polarization signals are then
combined with independent reference signals $E_{\rm ref\,1}$ and
$E_{\rm ref\,2}$ using $180^{\circ}$ hybrids. The four signals thus
formed, $(E_{L}\pm E_{\rm ref\,1})/\sqrt{2}$ and
$(E_{R}\pm E_{\rm ref\,2}) /\sqrt{2}$, are then amplified and
filtered by independent, phase- and gain-matched, chains of components.

After further gain and band-pass definition filters, each signal is
split in two, and sent to the radiometer and polarimeter branches. The
radiometer and polarimeter functions are logically distinct, and are
described separately below.

\subsection{Radiometer design} \label{sec:radiometer_design}

The continuous-comparison radiometer architecture of
\citet{Seiffert:2002bw} reduces $1/f$ noise from receiver gain fluctuations, but without the loss
of effective integration time of Dicke switch radiometers \citep{1946RScI...17..268D} and does so irrespective of the actual
knee frequency (the frequency at which the $1/f$ noise and the thermal
noise are equal). 
Considering only the right circularly polarized (RCP) channel, the sky signal voltage $E_{R}$ and
the
reference signal $E_{\rm ref}$ are combined using a $180\dg$ hybrid
to form sum and difference voltages $(E_{R} \pm E_{\rm ref})/\sqrt
2$ (where we chose units such that $E^2 = T$, the antenna
temperature). These voltages are amplified using identical gain chains
with power gain $G$. The two signals are then multiplied and
integrated. A change in gain $\Delta G$ will
cause the output to vary by \citep{Rohlfs:2004}
\begin{align}
\frac{\Delta T}{T_{\rm sys}} =\frac{\Delta G}{G} \frac{T_{R}-T_{\rm ref}}{T_{\rm sys}}.
\end{align}
i.e. if $T_{R}=T_{\rm ref}$, then $1/f$ gain fluctuations will have no effect on the measured
output. This comes at a price however: the level of thermal noise is $\sqrt{2}$
higher in a continuous-comparison radiometer than in a basic
radiometer because of an additional amplifier and the differencing of $T_{R}$
with the reference signal $T_{\rm ref}$.

The multiplier in the radiometer is implemented using a
second $180^{\circ}$ hybrid with detector diodes on each output. The
signals from the diodes are sampled, integrated, and subtracted in software to
obtain the multiplication product. An advantage of
this method is that we have access to the sky and load powers $T_R$
and $T_{\rm ref}$ individually in software. This is important for calibration
and diagnostics.

For the C-BASS instrument we implement a continuous-comparison radiometer
for each of the two orthogonal circular polarizations. The reference
signals are provided by two temperature-stabilized thermal loads.  In
terms of Stokes parameters, the powers in the left and right orthogonal
circular polarizations are $(I+V)/2$ and $(I-V)/2$ respectively, so the sum of the two
measured powers is proportional to Stokes $I$. The difference can
also be used to measure Stokes $V$.  Since the load temperatures
$T_{\rm ref\,1}$ and $T_{\rm ref\,2}$ are stable (see Section~\ref{sec:reference_loads}), variations
in the final measured quantity, $I -
(T_{\rm ref\,1} + T_{\rm ref\,2})/2$, represent the true sky brightness variations. 

Some continuous-comparison receivers (e.g., WMAP, \citealt{Jarosik:2003ie}) use a
second feed, pointed at a different part of the sky, as the
reference. This has the advantage that the receiver is well balanced,
i.e. the signals being correlated have very similar temperatures, as
the atmosphere (not relevant to WMAP) and CMB contributions to the sky temperature will be
common to both horns. However, using a second horn to provide the reference load is
unsuitable for the C-BASS experiment, which aims to measure the diffuse emission on large
angular scales. Any signal common to both horns, i.e. sky structure on
angular scales larger than the separation of the horn beams, is lost because a
continuous-comparison receiver measures the difference in horn powers. If the second feed
were feeding the same telescope optics, the maximum separation of the
beams would be only a few beamwidths, and all information on scales larger than a few degrees
would be lost. If instead we used a second feed-horn pointed at a large angle to the
telescope axis, the large beam of the reference feed-horn compared to the full telescope optics
would introduce ground contamination.

The reference load can also be a
thermal source at approximately the same physical temperature as the
antenna temperature (see e.g., {\em Planck} LFI \citealt{PlanckLFI}). A matched load, or
termination, at about the same
physical temperature as the sky signal can provide such a reference
signal. It should be kept at a constant temperature to ensure that any
variation in the output is due to the sky signal varying, and not the
reference load. The C-BASS reference loads are two matched, stabilized loads
housed in the receiver cryostat. These are discussed in more detail in
Section~\ref{sec:reference_loads}. The temperature of the loads can be set
to give optimum suppression of gain fluctuations, and is measured
by an accurately calibrated thermistor that can also be used to help
establish the temperature scale of the instrument.

\subsection{Polarimeter design} \label{sec:polarimeter}

Correlation polarimeters, in which the orthogonal electric field modes
are correlated, are the standard polarimeter architecture at radio wavelengths,
e.g., \citet{2011ExA....30...23B}. 
Defining correlation as $X*Y \equiv \avg{XY^*}$, we can obtain the
Stokes parameters thus \citep{Born:1964un}:
\begin{eqnarray}
I &=& E_L*E_L + E_R*E_R \nonumber \\
Q + iU &=& 2E_R*E_L \nonumber \\
V &=& E_L*E_L - E_R*E_R,
\end{eqnarray}
i.e., the linear polarized Stokes parameters $Q$ and $U$ are the real
and imaginary parts of the correlation between left and right circular
polarizations. Receivers that implement the correlation approach are
immune to amplifier $1/f$ noise as it is uncorrelated between amplifiers. Additive noise (e.g. the
thermal noise from the amplifier) and correlated noise (e.g. due to common physical temperature
changes of the amplifiers) are not suppressed. 

In the C-BASS polarimeter, the first stage is to recover the pure
circular polarization signals from the cryostat outputs, which are
linear combinations of the right and left circular polarizations with
their respective load signals. As the polarization measurement is immune
to $1/f$ noise, the reference load signals are
unnecessary and would add excess noise to it. The combined sky and load
signals are each passed through a
second $180^{\circ}$ hybrid, where the reference signals $E_{\rm
  ref\,1}$ and $E_{\rm ref\,2}$ are discarded. The recovered circular
polarizations $E_{L}$ and $E_{R}$ are phase switched and correlated to
obtain two measurements each of Stokes $Q$ and $U$. The multiplying
elements of the correlator are implemented using hybrids to combine
the signals, and detector diodes to form the squares of these sums, as
in the radiometer. The complex products are formed as follows. The
outputs of the $180^{\circ}$ hybrids are proportional to
$$
E_L+E_R, E_L-E_R
$$
and so the outputs from the detector diodes are
$$
\lan E_L^2 \ran + \lan E_R^2\ran + 2\lan E_LE_R\ran, \lan E_L^2\ran + \lan E_R^2\ran - 2\lan
E_LE_R\ran.
$$ 
Taking the difference between these outputs gives a signal
proportional to the product $\lan E_LE_R\ran$, which is the real part of the
complex correlation, $Q = 2\Re \lan E_L E_R\ran$. Similarly, using $90^{\circ}$
hybrids, the differenced output of the diodes is proportional to
$\lan iE_LE_R \ran$, which gives the imaginary part, $U = -2\Im \lan E_L E_R\ran$. Any
residual component of the total power terms $\lan E_L^2 \ran$ and $\lan E_R^2\ran$ that
is not perfectly removed, due to imbalances in the hybrids and post-hybrid hardware, is
removed by the phase switching described in Section~\ref{sec:phase_sw}.

The detector output voltages are routed to the digital readout system
that performs further signal processing, described in
Section~\ref{sec:detectors_and_digital_backend}. The final data product is read from the
digital readout system by a data acquisition computer and stored to
disk.

\subsection{Phase Switching}\label{sec:phase_sw}

In both the radiometer and the polarimeter, phase switches are used to
invert the phases of the signals prior to the correlating
element. Phase switching effectively swaps the outputs of each hybrid,
resulting in the sky and load signals at the output of the differencing
being switched in sign. Phase switches (see Section~\ref{sec:phase_sw:hardware}) are placed in both
arms of each pair of signals, and are driven by digital IO pins on the backend FPGA (see
Section~\ref{sec:detectors_and_digital_backend}). One phase switch in each pair is switched by a
500\,Hz square 
wave signal, while the second is driven by another 500\,Hz square wave offset from the first by
$90^{\circ}$ of phase. The switches are cycled between all four combinations of states at an
effective phase switch rate of 1\,kHz, with each sample of the integrated output signal containing
data from equal times in each state. This ensures that any gain differences and offsets between the
physical channels are differenced out in the integrated data. 
The phase switch frequency was selected to be higher than the 60\,Hz mains signal, while still being
low enough to avoid significant attenuation of the high-order harmonics of the phase switch signal
by the video bandwidth (VBW) of the detector diodes.

\begin{figure}
 \centering
 \includegraphics[width=8cm]{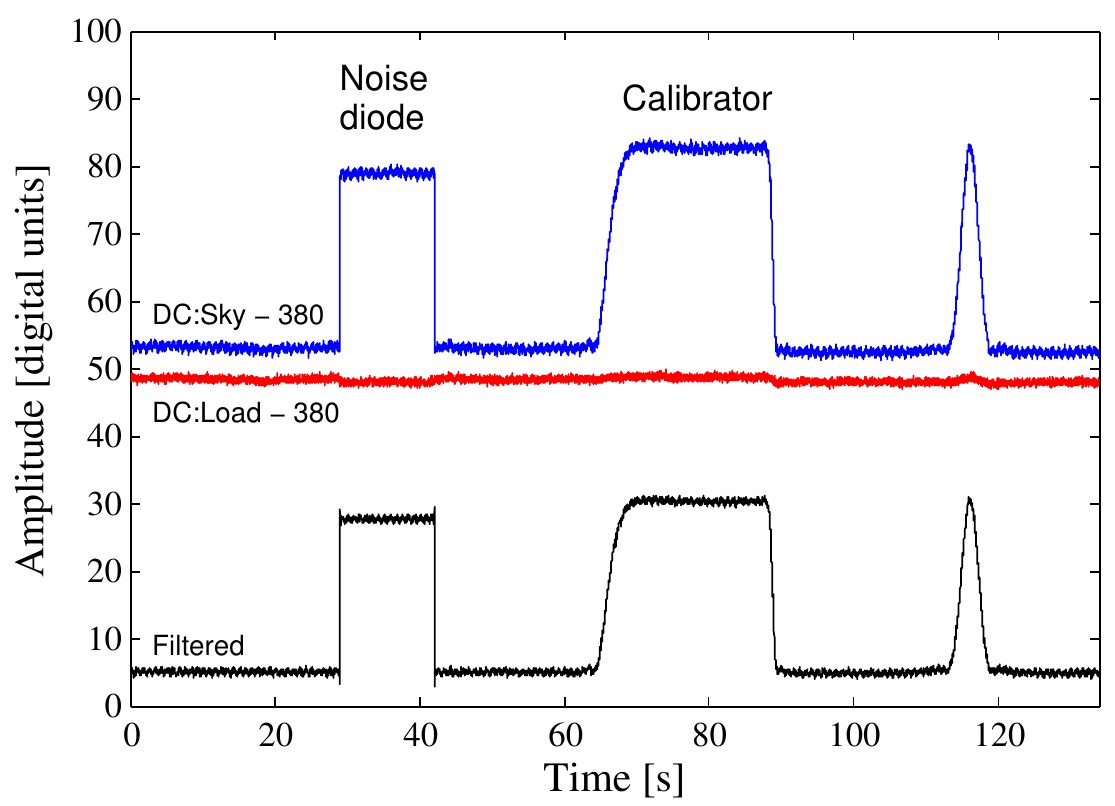}
 \caption{Total intensity data from the filtered and DC-coupled data streams described in
Section~\ref{sec:phase_sw}. In this stretch of data the noise diode is fired, then the
telescope is pointed at a calibrator source. The blue trace shows data from the phase switch state
in the filtered data stream that produces the sky signal at the diode output, while the red trace
shows the phase switch state that produces the load signal. These traces have been offset as
indicated for clarity. The black trace is the corresponding section of the filtered data stream,
which is a filtered version of the difference between the blue and red traces.}
 \label{fig:data_streams}
\end{figure}

Since phase switching moves the signal of interest from the
low-frequency part of the spectrum ($\lesssim 20\,$Hz) to kHz
frequencies, it is possible to filter out lower frequency
contaminating signals, such as $60\,$Hz mains pickup, before demodulating
and recovering the sky signal. While it is possible to do this high-pass
filtering in hardware, this has the effect of removing the
absolute level of the diode output, and thus losing important
diagnostic information about the state of the receiver. Instead, we
directly sample the DC-coupled voltages from the detector diodes, and
perform two parallel reductions of the data in real-time firmware
(known as DC-coupled and filtered modes). In
the DC-coupled mode, the data are simply averaged in to corresponding phase switch
states for each diode. This means that the sky and load signals are
available independently, providing diagnostic information about the
power levels, but containing some contamination, particularly from
mains pickup. In filtered mode, the data are high-pass filtered,
demodulated and differenced before averaging. This loses information
about the individual diode signals, as only the difference (sky $-$
load) is stored, but provides the best quality
science data with 60\,Hz pickup and other potential low-frequency contaminant signals removed. A
short period of total intensity data from these data streams is plotted in
Fig.~\ref{fig:data_streams}, illustrating the relationship between the two data streams. Both data
streams (DC-coupled and filtered) are recorded by the readout computer
(Section~\ref{sec:detectors_and_digital_backend}).

\subsection{Noise diode calibration}

A noise diode is used to calibrate both the polarimeter and
radiometer. A broadband, temperature-stabilized noise diode signal is
split and coupled in to the right and left circular polarization
signals in the cryostat using $-30\,$dB directional couplers. The coupled
noise power is equivalent to an antenna temperature of about $3\,$K and is
therefore detected with a large signal-to-noise ratio of $>200:1$ in a 10\,ms integration, but
without changing the total system power by more than 10\%. The noise
diode can be switched on and off under control of the digital backend. 
The noise signal is coupled in to the sky signal path and should appear purely in
the sky signal data channel. 
A detection of the noise diode signal in the load
data channel is an indication of amplitude or phase imbalance between the
first and second hybrids. The typical level of this leakage can be seen in the data plotted in
Fig.~\ref{fig:data_streams}.
We use this as a diagnostic only, as the
full phase switching regime automatically cancels out the effect of
such imbalances. The amplitude of the noise diode signal is used to track gain variations
in the system between astronomical calibration events.

In the polarimeter, the noise diode should generate a pure $Q$ signal
in the instrument reference frame,
since it is fully correlated between the two circular
polarizations. Presence of the diode signal in $U$ indicates phase
imbalance between the circular polarizations. This is monitored by
regularly firing the noise diode and the effect is calibrated out of the
data. The noise diode is typically fired on minute time-scales to
track receiver gain fluctuations. The amplitude of the noise diode
signal is calibrated against astronomical sources of known (and
stable) flux density on time-scales of hours.  A consequence of the
noise diode injection system is that there is a small correlated noise
signal injected even when the diode is switched off, due to thermal
noise in the diode, and this produces a small offset in the $Q$
channels of $\approx 30\,$mK that is removed in subsequent calibration.

\section{Cryogenic receiver} \label{sec:implementation_cryostat}

\subsection{Cryostat design}

\begin{figure}
 \centering
 \includegraphics[width=8cm]{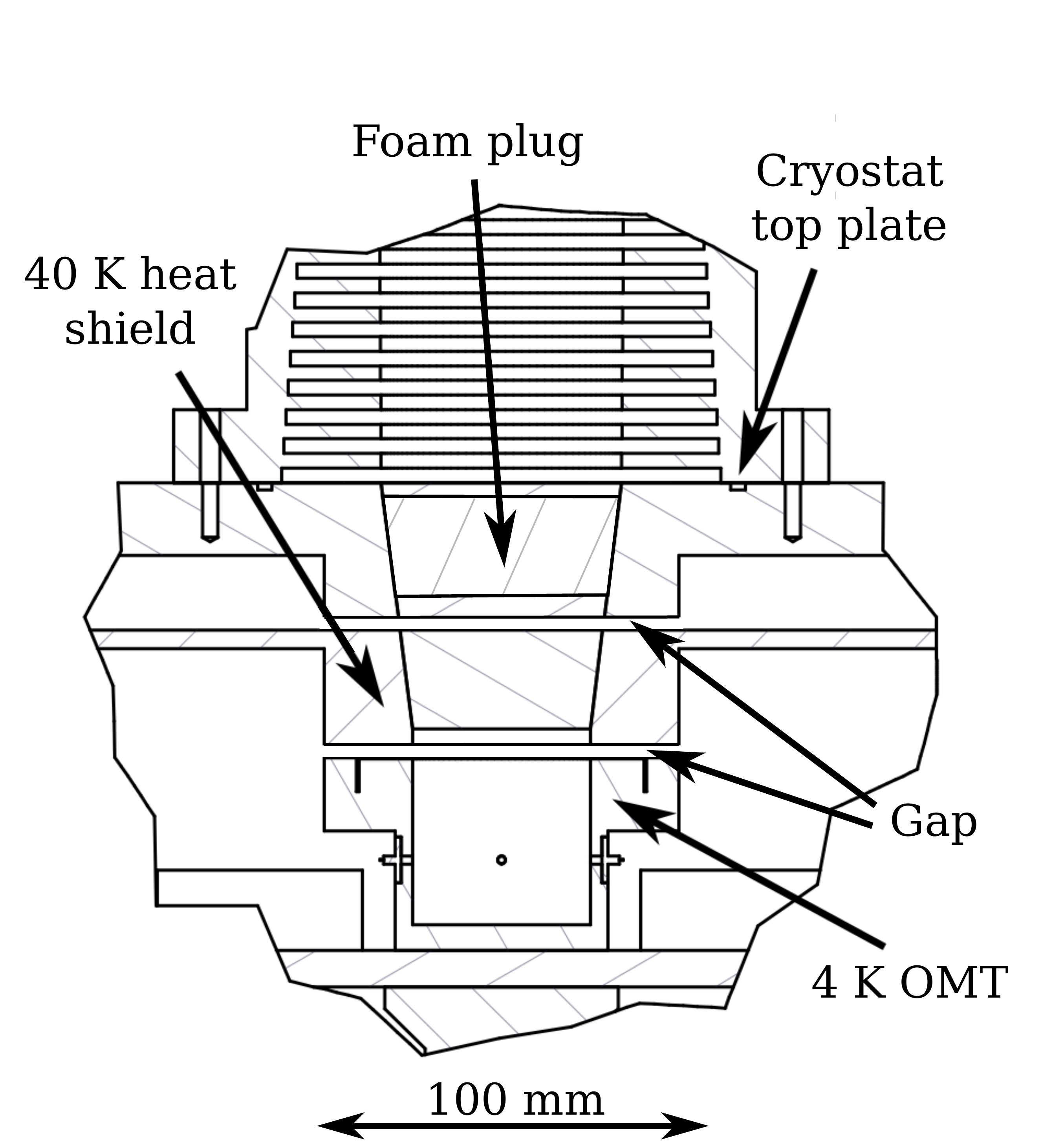}
\caption{A cross-section of the optical feedthrough on the
  cryostat. The smoothly-tapered section of the horn throat is
  incorporated in to the top wall of the cryostat and the $40\,$K heat
  shield. A Mylar sheet (not shown) between the corrugated horn
  section and the cryostat body provides a vacuum window. A $0.5\,$mm
  vacuum gap (exaggerated in the drawing) separates the $300\,$K stage and the $40\,$K thermal
  shield. A similar break separates the $40\,$K heat shield and the
  $4\,$K OMT.}
 \label{fig:horn_to_OMT_cryostat transition}
\end{figure}

The cryostat cools the low-noise amplifiers (LNAs) and pre-LNA
components to reduce the system temperature, and to provide a cold
load at comparable temperature to the sky. A 2-stage Sumitomo
SRDK-408D2 Gifford-McMahon cryocooler driven by a CSA-71A air-cooled
compressor is used to provide cooling. 
The cryostat is cylindrical, coaxial with the feed-horn. The
corrugated parts of the feed-horn are bolted directly to the cryostat
top plate, while the smoothly-tapered horn throat sections are
machined in to the cryostat body and the internal $40\,$K heat
shield (see Fig.~\ref{fig:horn_to_OMT_cryostat transition}). Gaps of
$0.5\,$mm between the stages provide thermal isolation without
compromising radio frequency (RF) performance. A Mylar window provides vacuum isolation,
and a plug of Plastazote LD45 foam glued in to the 300$\,$K smooth horn
section provides infra-red blocking.

A circular copper cold plate is mounted directly to the $4\,$K cold
head. All the $4\,$K components are mounted on this plate. Stainless
steel coaxial cables are used to transfer the RF signals from the LNA
outputs to the amplification and notch filtering (Section~\ref{sec:notch_filters}) on the $40\,$K
stage, and then from the $40\,$K stage to the $300\,$K outer cryostat wall.

\subsection{Orthomode transducer}

\begin{figure}
 \centering
 \includegraphics[width=8cm]{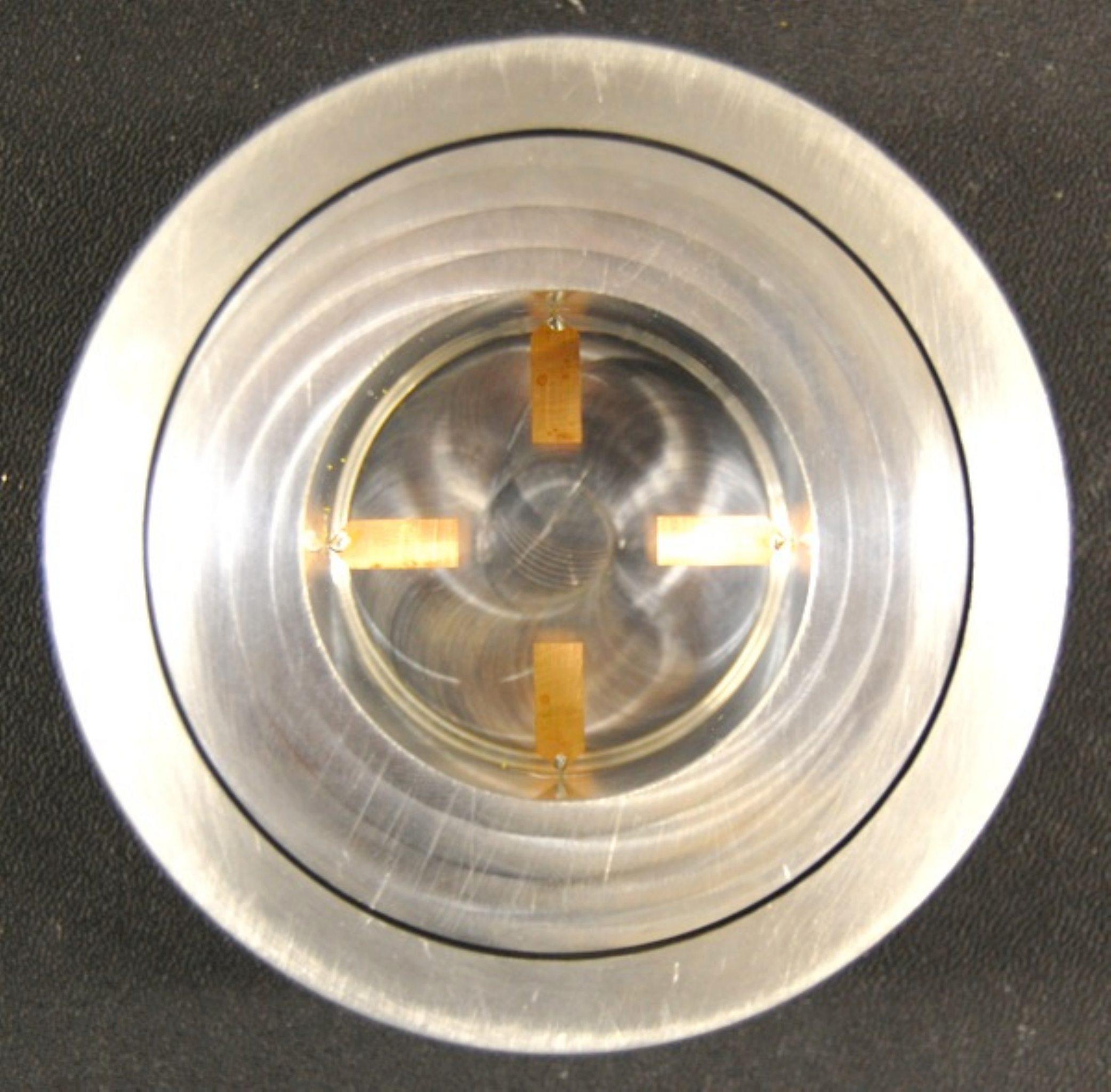}
 \caption{A top view of the C-BASS OMT, showing four rectangular probes suspended orthogonally in the waveguide.}
 \label{fig:OMT_picture}
\end{figure}

The orthomode transducer (OMT), which extracts orthogonal linear
polarization voltages from circular waveguide, is attached to the $4\,$K
stage in the cryostat. It must thus be compact and easily coolable to
$4\,$K; these conditions are not met by most commercial OMTs. 
We developed an OMT based on a design by
\citet{Bock:1999b}, comprising four probes at right angles in a
cylindrical waveguide \citep{Grimes:2006ef}. The OMT body is machined
from aluminium and contains four rectangular probes, chemically etched
from a 0.1\,mm thick copper sheet, placed orthogonally in a single
cross-sectional plane of a circular waveguide. The probes are
supported only by the pins of the SMA connectors that protrude through
the waveguide wall, as shown in Fig.~\ref{fig:OMT_picture}. There is a narrow ($0.5\,$mm) break in
the waveguide between the top of the OMT and the start of the horn, which is incorporated in to the
$40\,$K heat shield. Simulations conducted in Ansoft's HFSS
software\footnote{\url{http://www.ansoft.com/products/hf/hfss/}} suggest that the effect of the
break on the cross-polarization of the OMT is negligible, increasing it from $-76\,$dB to $-75\,$dB at
5\,GHz.

The bases of the probes are tapered at an angle of 45$^{\circ}$ to the waveguide
walls to reduce the capacitative coupling to the waveguide wall and are soldered to the pins of the
SMA jacks, which are grounded to the body of the waveguide and fed
through the wall in front of a fixed backshort. The OMT has two
outputs for each linear polarization, which must be combined
externally in anti-phase in the linear-to-circular converter (see Section~\ref{sec:L2C})  to give the
full output signal. The signals from each pair
of probes are transferred by equal-length, semi-rigid cables to the
linear-to-circular converter to be combined to produce the orthogonal
circular polarization signals.

\subsection{Cold loads and temperature stabilization} \label{sec:reference_loads}

The reference loads are shunt resistors attached to a copper
bobbin that is mounted on the top of the cold plate. A heater coil is wound
round the bobbin and a Cernox CX-1050-AA temperature sensor is embedded in it. The
bobbin is mechanically bolted to the $4\,$K cold plate, separated from
it by Teflon washers to ensure a weak thermal link. The
thermal link is strong enough to cool the bobbin (giving the heater
something to work against), but weak enough that the temperature of
the bobbin does not greatly affect the cold plate temperature. The
bobbin can be heated to $80\,$K without raising the cold plate
temperature by more than $1\,$K. The reference loads are connected to the
$180\dg$ hybrids through stainless steel coaxial cables, which also
minimize the thermal coupling between the bobbin and the cold plate.

\begin{figure}
 \centering
\includegraphics[width=8cm]{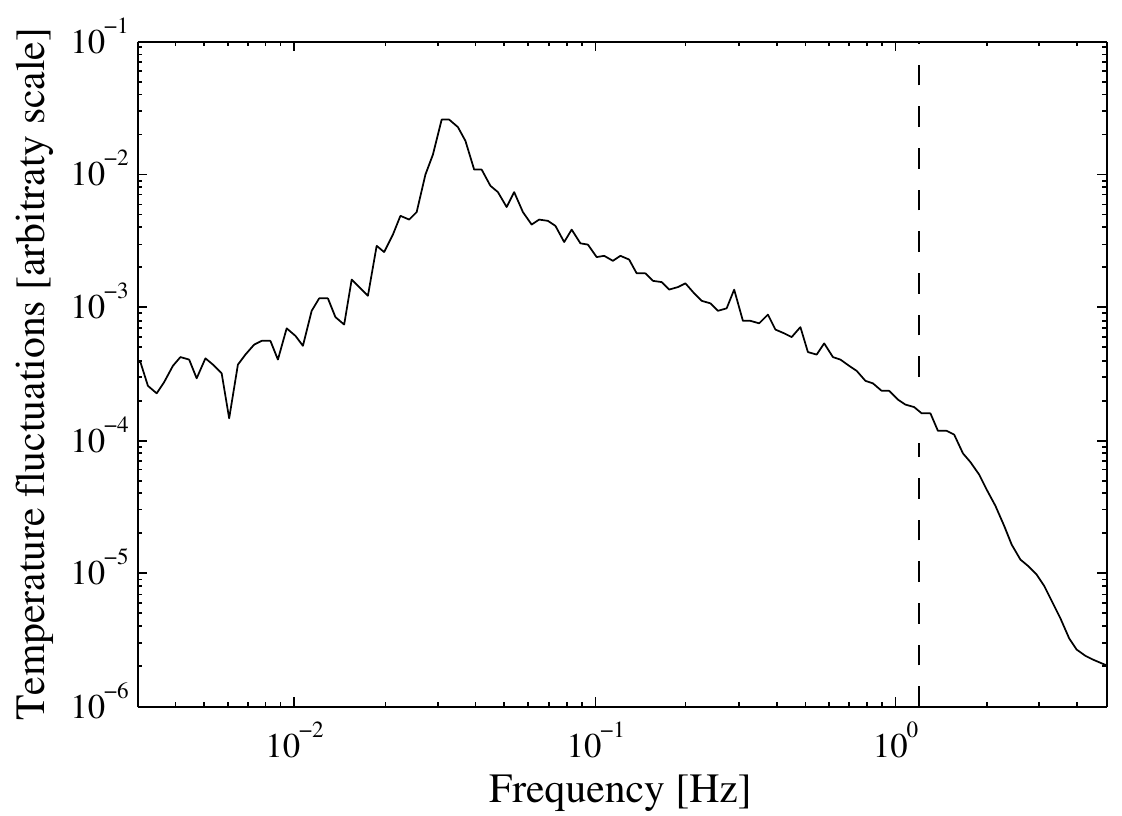}
\caption{
Power spectrum of the cold load temperature sensor data. Note the low level of
  mechanical refrigeration cycle signal at $1.2\,$Hz (indicated by the dashed line). The fluctuations
  have an rms of $\approx 0.4\,$mK.}
 \label{fig:cold_load_stability}
\end{figure}

A Cryocon 32B temperature controller, operating as a proportional-integral-derivative (PID) control
loop, is used to keep the bobbin temperature stable. A major potential
contaminant of the receiver data stream is the thermal variation due
to the refrigeration cycle, which varies with a fundamental frequency
of $1.2\,$Hz. The measured stability of the cold load temperature sensor data is shown in
Fig.~\ref{fig:cold_load_stability}. The refrigeration cycle is
not visible in the temperature power spectrum, and the
temperature is stable with an rms value of $\approx 0.4\,$mK.

\begin{figure}
 \centering
\includegraphics[width=8cm]{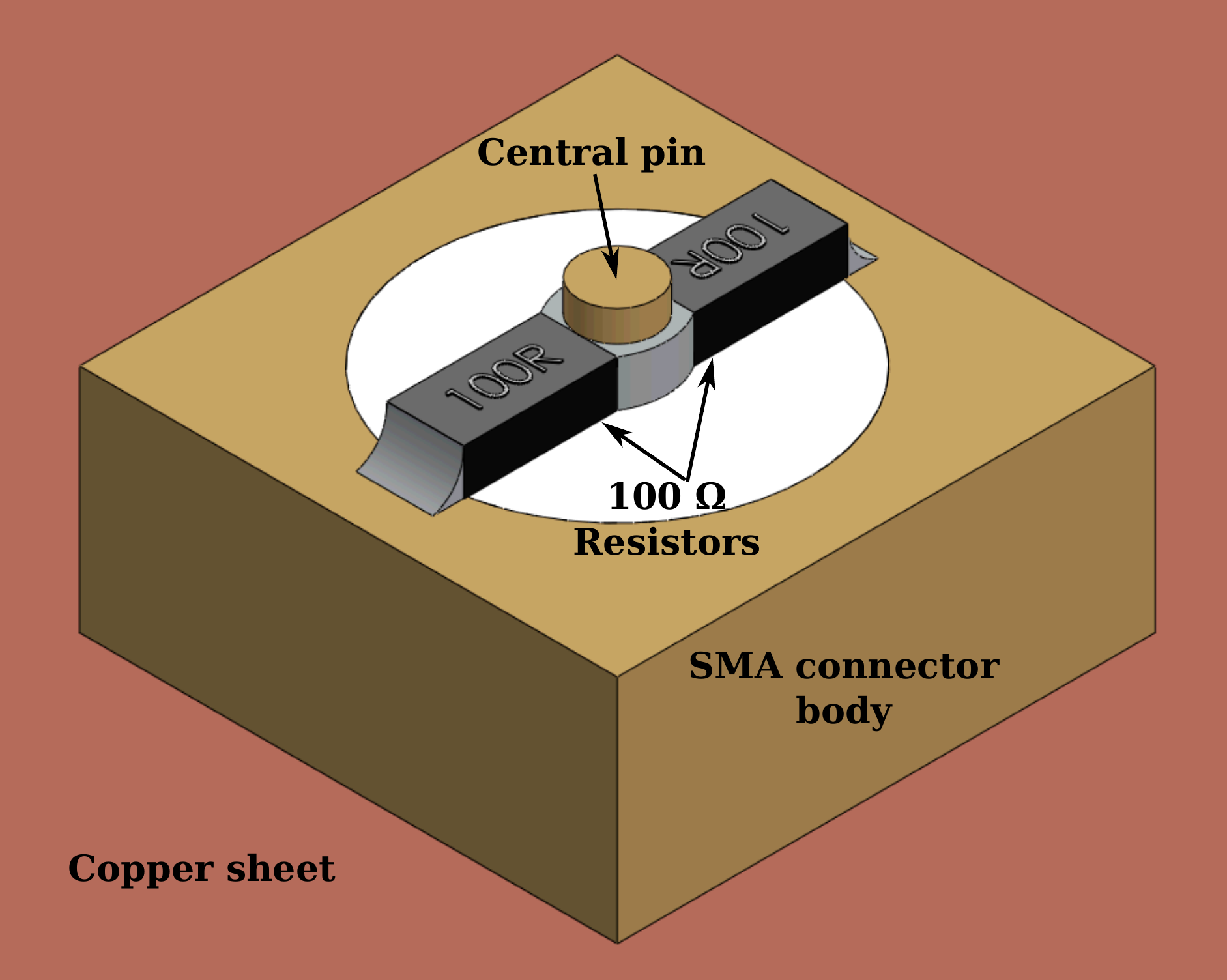}
\caption{The C-BASS cold reference loads, rear view. An SMA straight
  PCB receptacle is soldered in to a copper sheet that is attached to a
  copper bobbin, whose temperature is controlled. Two shunt
  $100\,\Omega$ resistors are soldered between the central pin and
  outer conductor of the coaxial line. This design ensures that the
  resistive element of the termination is at the same physical
  temperature as the copper bobbin.}
 \label{fig:cold_load_assembly}
\end{figure}

Initially we used commercial $50\,\Omega$ SMA RF terminations, soldered
in to pockets in the copper bobbin, as the reference loads. However,
the mechanical construction of the terminations meant that the shunt
resistors embedded therein were weakly thermally linked to the outer
body of the termination, resulting in an uncertain load
temperature. We developed a second design based on SMA bulkhead connectors,
in which the thermal link is
stronger ensuring that the resistors are at the same temperature as the
copper bobbin. In this new design two surface-mount
$100\,\Omega$ resistors are soldered between the centre pin and body of
the connector. This is then mounted on a small copper plate, which is
in turn mounted to the top of the bobbin. A 3D CAD model of this is shown in
Fig.~\ref{fig:cold_load_assembly}. This design ensures that the
resistors (which are poor thermal conductors) are well thermally
connected to the bobbin. S-parameter tests with a vector network
analyser confirmed that the resistors gave a good match to $50\,\Omega$
coax at cryogenic temperatures.

\subsection{Hybrids}

The receiver contains both $180\dg$ hybrids (that combine two
input signals with 0 and $180\dg$ phase shifts) and $90\dg$ hybrids
(that combine two input signals with $90\dg$ phase shifts). Both $90^{\circ}$ and $180^{\circ}$
hybrids are used before the LNAs in the signal chain; their
loss is thus of critical importance. No commercial products had
sufficiently low loss and good performance over the C-BASS band, so we
designed our own hybrids.

\subsubsection{$180\dg$ Hybrids} \label{sec:180hybrid}

\begin{figure}
 \centering
\subfloat[]{
 \includegraphics[width=0.35\textwidth]{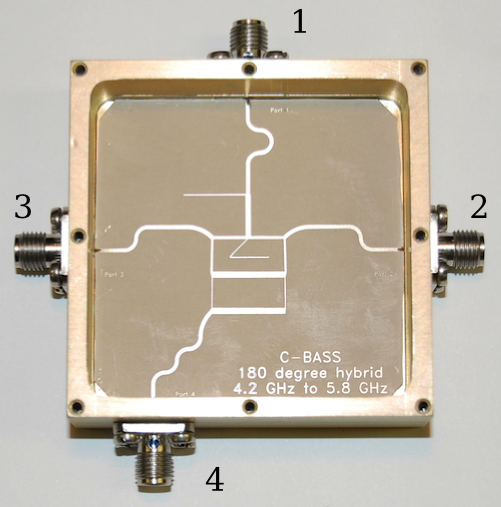}
}\\
\subfloat[]{
 \includegraphics[width=0.46\textwidth]{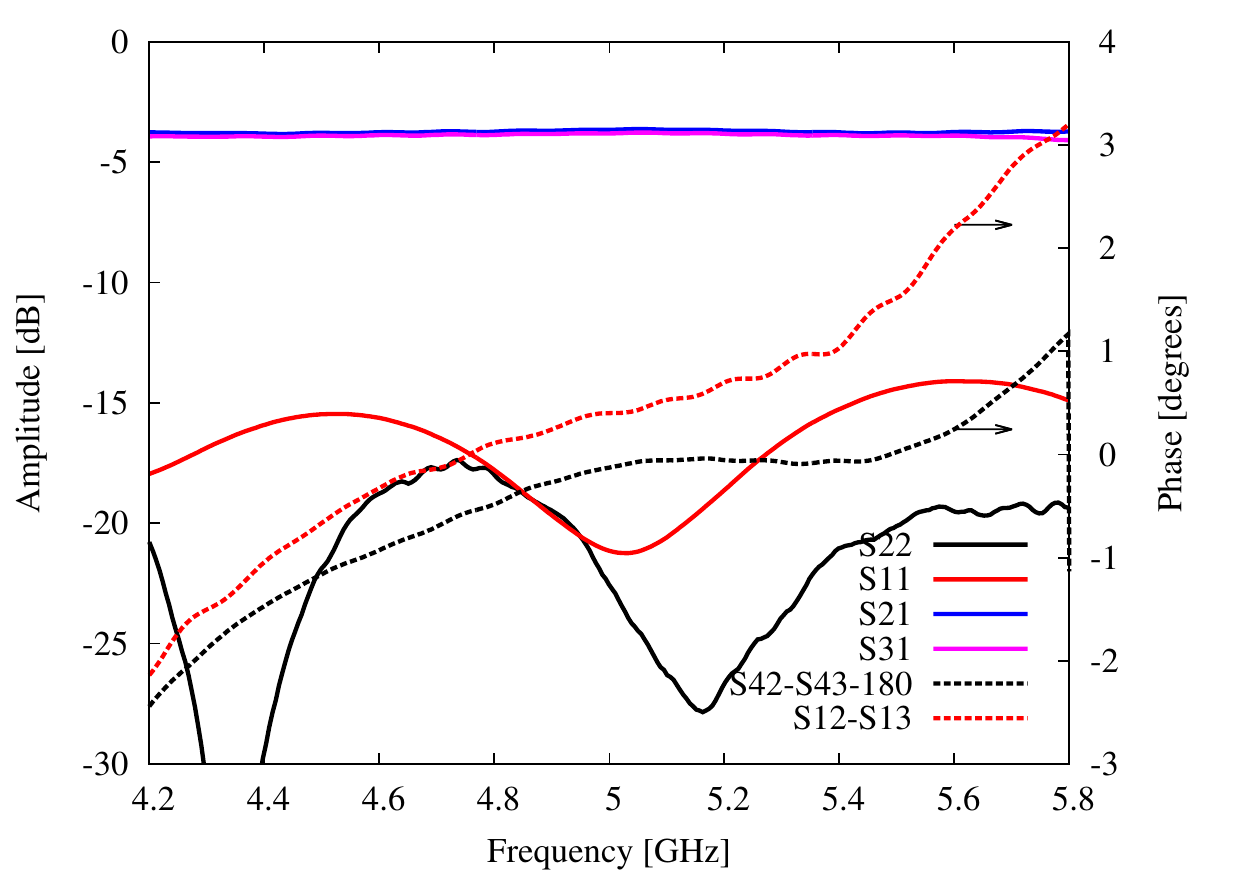}
}
\caption{(a) A picture of a C-BASS $180\dg$ hybrid. Ports 2 and 3 are
  the input ports. Port 1 produces the sum of the voltages at ports 2
  and 3, and port 4 produces the difference. The box housing the substrate is 50\,mm square. (b) The
measured response
  of the $180\dg$ hybrid, showing the return loss (S11 and S22), the
  transmission to the sum port (S21 and S31), and the phase of the sum
  ($\angle S12-\angle S13$) and difference ($\angle S42-\angle S43-180^{\circ}$) operations. The
phase shifts are within $1.\!\dg5$ of ideal over the full C-BASS band of 4.5 to
  $5.5\,$GHz. Amplitudes are plotted against the left-hand ordinate
  and phases (indicated by the arrows) against the right-hand ordinate.}
 \label{fig:180hybrid}
\end{figure}

Achieving a broadband $180\dg$ phase shift while keeping the loss low
is a challenging engineering problem. The C-BASS $180\dg$ hybrids are
based on the `coat hanger' model \citep{Knochel:1990}, which is an
extension of the rat-race hybrid \citep{Pozar:2005} (the name `coat
hanger' comes from the characteristic shape of the copper traces on the substrate).
The hybrid is implemented on low-loss Rogers Duroid 6010 substrate. 
A picture of one of the C-BASS $180\dg$ hybrids, and a
plot of the measured response, is shown in
Fig.~\ref{fig:180hybrid}. 

The performance of the hybrid shown in Fig.~\ref{fig:180hybrid}(b) is close to ideal: the
transmission from the
inputs (ports 2 and 3) to the sum port are equal to within $0.1\,$dB
over the C-BASS band. The return loss of input port 2 (S22) is typical
of all the ports, and is better than $-15\,$dB. The phase performance
is particularly noteworthy, coming within $1.\!\dg5$ of ideal operation.

\subsubsection{$90\dg$ Hybrids} 

The receiver uses $90^{\circ}$ hybrids in the polarimetry section to combine
the circular polarizations with a $90^{\circ}$ phase shift, and as
circularizers in the linear-to-circular converter described 
in Section~\ref{sec:L2C}, producing orthogonal
circular polarizations from the orthogonal linear polarizations produced by
the OMT. Different designs were implemented in each case.

\begin{figure}
 \centering
\subfloat[]
{
 \includegraphics[width=0.35\textwidth]{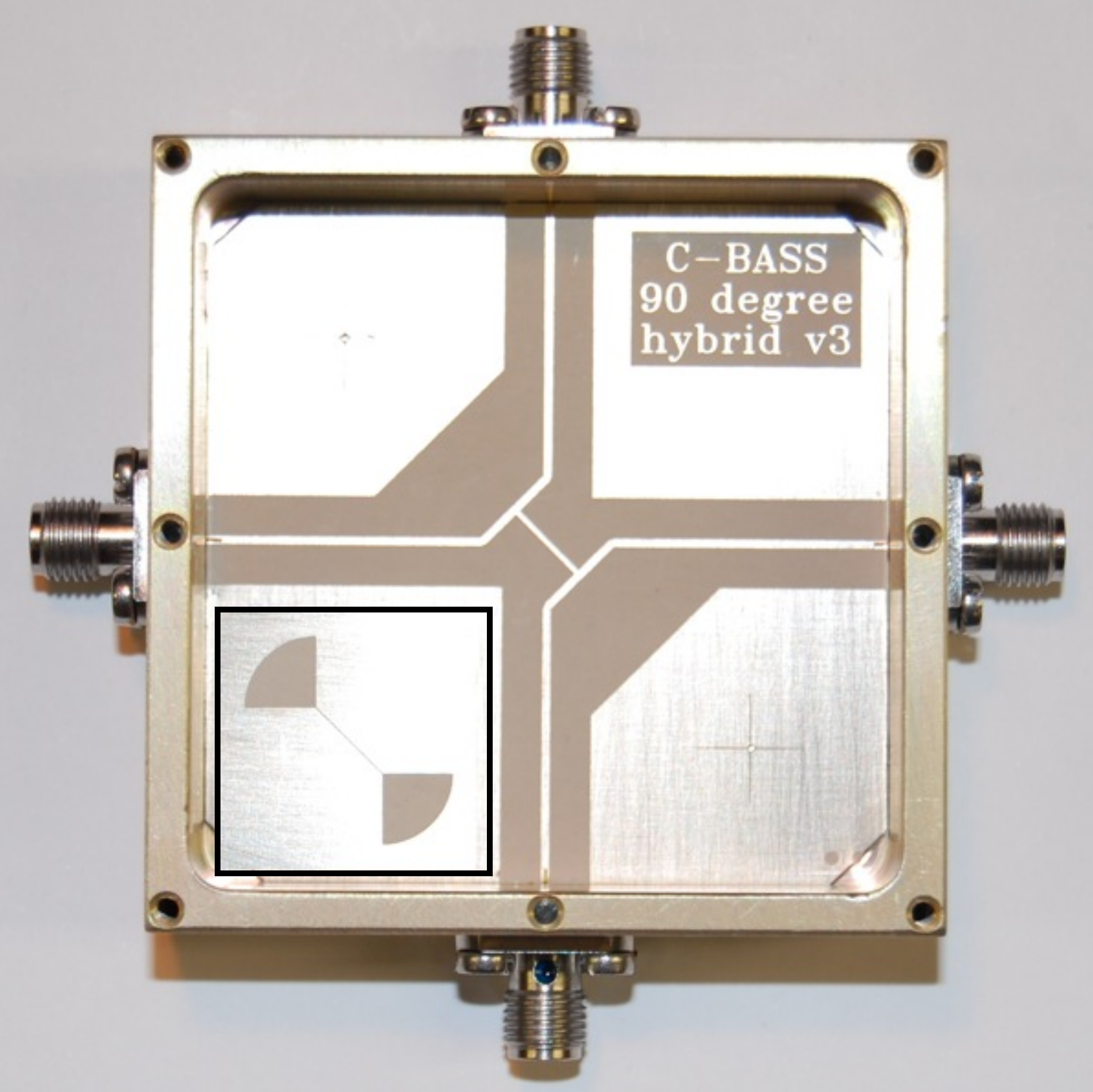}
}\\
\subfloat[]
{
 \includegraphics[width=0.46\textwidth]{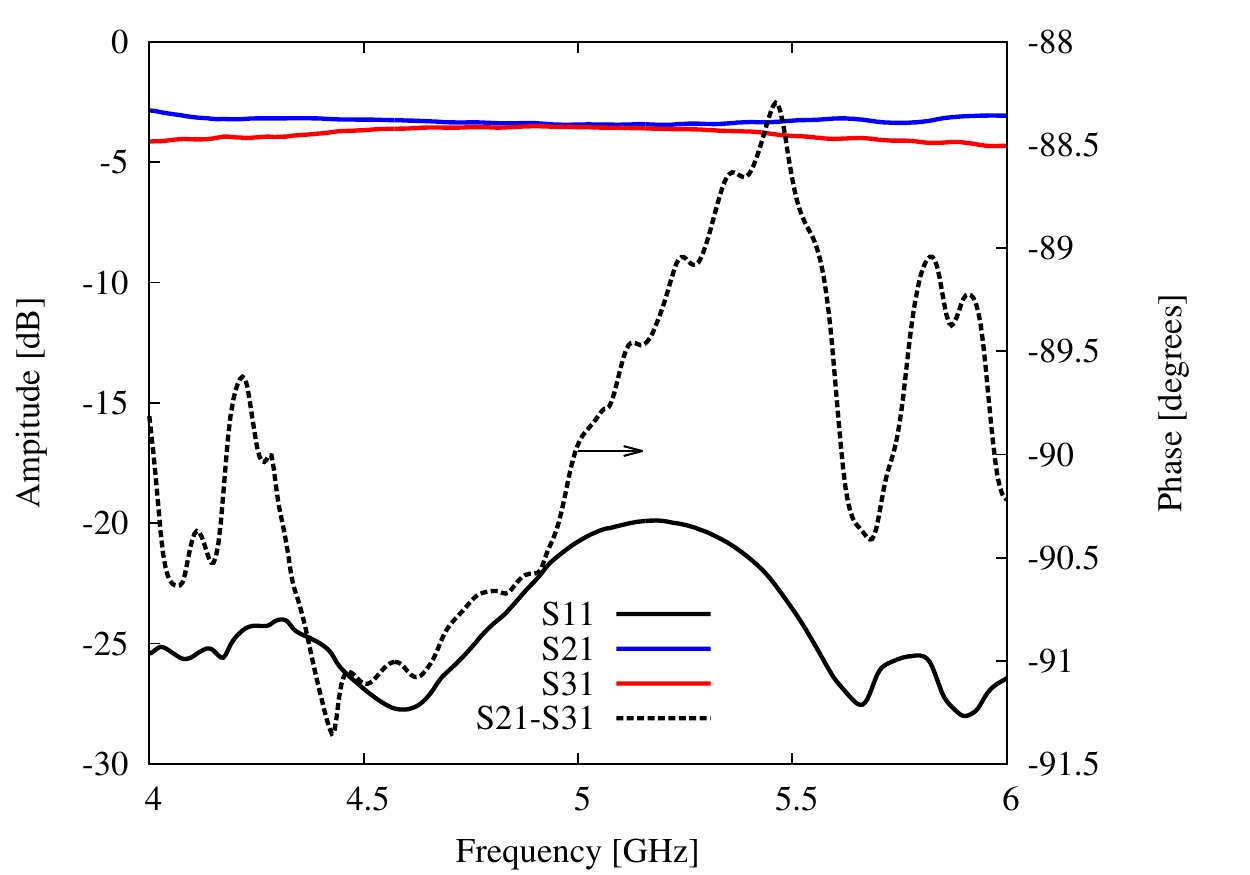}
}
\caption{(a) Picture of one of the microslot $90^{\circ}$ hybrids with the
  ground plane slot line shown as an inset. The slot lies directly
  underneath the central connecting microstrip line. The box housing the substrate is 50\,mm square.
(b) Measured response of a typical microslot $90^{\circ}$ hybrid. The
phase (indicated by the arrow) is plotted against the right-hand ordinate.}
\label{fig:microslot_hybrid}
\end{figure}

The $90\dg$ hybrids in the C-BASS polarimeter are based on the microslot
coupler design \citep{deRonde:1970,Hoffman:1982,Hoffman:1982b}. They
have nearly ideal phase response -- within $1\dg$ of a $90\dg$ phase shift
over the C-BASS band -- and a return loss better than $-20\,$dB, as
shown in Fig.~\ref{fig:microslot_hybrid}. The microslot in the
bottom ground plane of the circuit lies directly underneath the
central coupling line. This requires there to be a machined void in
the mounting box beneath the microslot. This design is too large to
fit in the space available for the linear-to-circular converter, so
a different hybrid design, a microstrip branch-line coupler, is used
instead.

\subsection{Linear-to-circular converter} \label{sec:L2C}

The out-of-phase signals from opposite pairs of probes in the OMT need
to be combined with a $180^{\circ}$ phase shift to obtain orthogonal
linear polarizations. The linear polarizations are then passed through
a $90^{\circ}$ hybrid to produce orthogonal circular polarizations. We
combine these functions in to a single planar circuit that we call a
linear-to-circular converter (L2C).

The signals from opposite pairs of probes are `subtracted' (combined with a $180^{\circ}$ phase
difference) with a 3-port
device based on the $180^{\circ}$ hybrid described in
Section~\ref{sec:180hybrid} that we call a `subtracter'. The signals from opposite probes are
connected
to ports 2 and 3 of the $180^{\circ}$ hybrid, and the output signal is
then seen at port 4. However, since the signals at ports 2 and 3 are
$180^{\circ}$ out of phase, port 1 is at virtual ground. Port 1 can
therefore be removed entirely without affecting the performance of the
device at combining $180^{\circ}$ out of phase signals. The two
`subtracters' and branch line $90^{\circ}$ coupler are integrated on
to a single substrate and boxed. This box is attached to the cold
plate close to the OMT to keep the cable length between the OMT and
L2C short (10\,cm).

\subsection{Low noise amplifiers}

The northern C-BASS receiver uses Low Noise Factory\footnote{\url{http://www.lownoisefactory.com/}} LNC4\_8A low noise
amplifiers (LNAs), which have excellent noise ($3.3 \pm 0.4\,$K) and gain ($41.6 \pm 0.3\,$dB)
performance between 4 and 6\,GHz at cryogenic temperatures. These are preceded in the signal
path by Raditek cryogenic isolators; while the isolators result in a small noise temperature penalty
of $< 1\,$K (typical insertion loss is $<0.1\,$dB), they improve the matching to the LNA.

Special attention was paid to the wiring of the bias supply. Improper
grounding of the wiring can result in detectable pickup of the
$60\,$Hz supply voltage in the receiver data. The cryostat body is used
as the return path for the DC bias currents. The final instrument
configuration shows extremely low levels of 60\,Hz pickup: it is below the noise floor of a power
spectrum of 20\,min of data, implying an equivalent temperature of $<30\,\mu$K.

Initially, eMERLIN C-Band LNAs were used in the cryostat. These amplifiers had typical noise
temperatures of
$\approx 12\,$K at $5\,$GHz with a gain of $\approx 30\,$dB. The effect of
pre-LNA losses and loading on the system by the absorbing baffles (see
\citealt{Holler:2012dp} and S.J.C. Muchovej et al. in prep), along with LNA stability issues, required us
to replace them with the Low Noise Factory amplifiers described here
in order to achieve a satisfactory system temperature.


\subsection{Notch filters} \label{sec:notch_filters}

\begin{figure}
 \centering
\subfloat[]{ 
\includegraphics[width=0.47\textwidth]{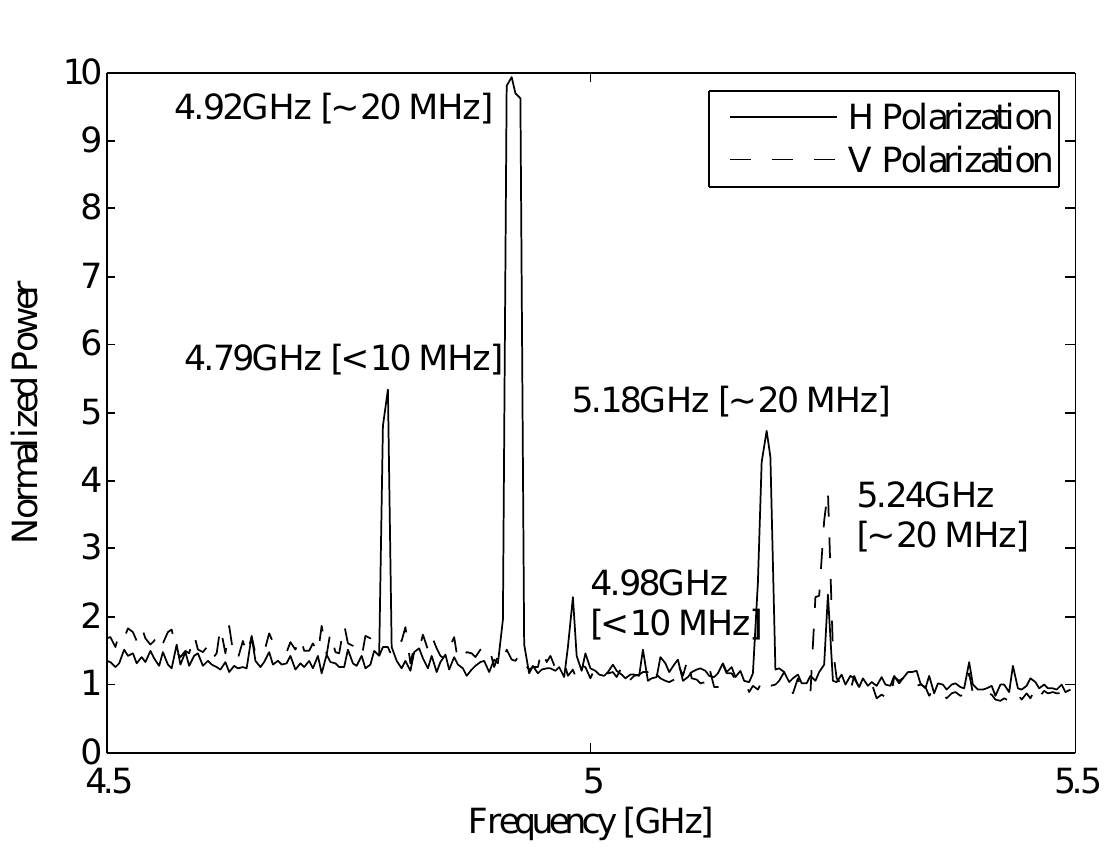}
 \label{fig:rfi_environment}
}\\
\subfloat[]
{ 
\includegraphics[width=0.47\textwidth]{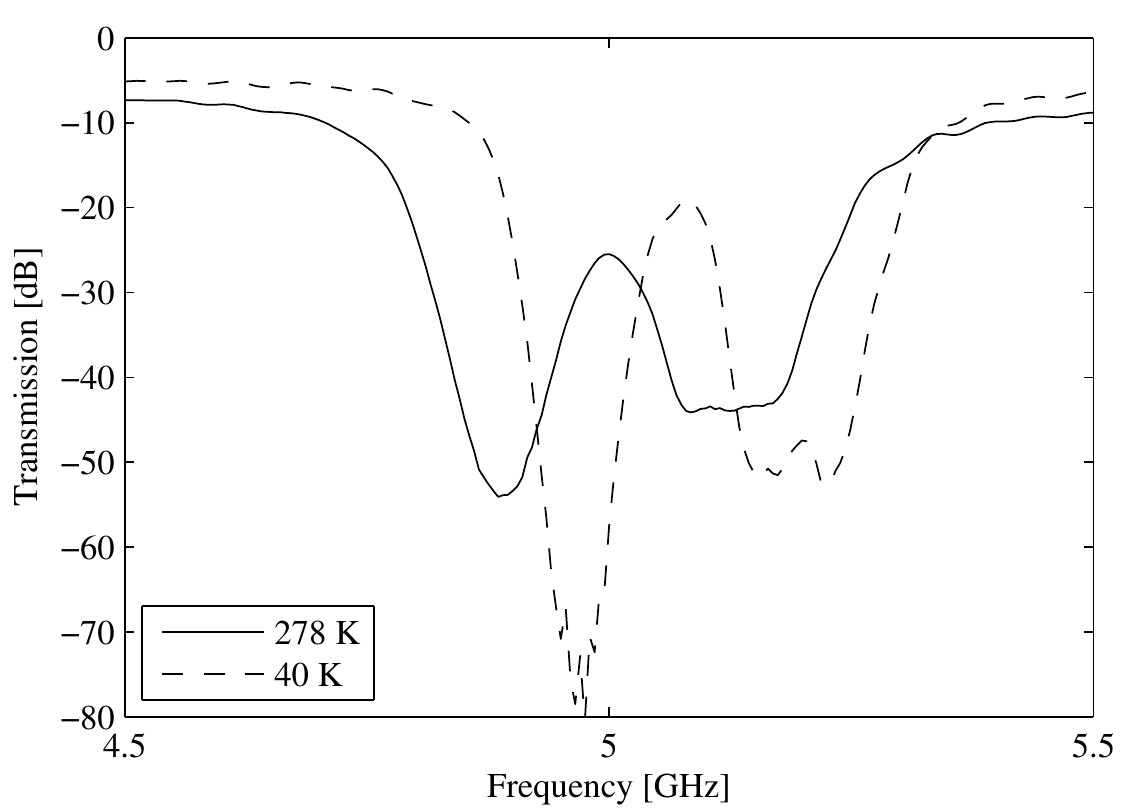}
 \label{fig:NotchFilterTempResponse}
}
\caption{(a) An unfiltered measurement of the terrestrial RFI environment as measured by
   a spectrum analyser attached to a C-band horn. Not shown
   are the strong satellite broadcast signals at $\approx 4\,$GHz. The spectrum is dominated by
microwave point-to-point transmitters located to the north and south of OVRO.
(b) The effect of cooling on the notch filter transmission;
   cooling the filter to $40\,$K increases the quality factor
   ($Q=\nu_{c}/\Delta \nu$) by reducing the resistive losses in
   the copper, and changes the notch centre frequencies. The depth of
   the notches in the $40\,$K measurement are limited by the noise
   floor of the measurement equipment. The higher $Q$-factor is seen
   in the increased slope of the notch edges.}
\end{figure}

Man-made radio frequency interference (RFI) has proven to be a
major challenge for the northern survey. Typical sources of RFI
include aircraft radar/transponders, geostationary C-Band broadcast
satellites, and fixed microwave point-to-point links (see
\fign{fig:rfi_environment} for a typical unfiltered spectrum). Aircraft
interference is sporadic enough that we can flag it in the data
reduction pipeline (typically a few percent of data are flagged with aggressive filtering). 
However, the permanent nature of satellite and
terrestrial RFI requires a different approach.

The C-Band satellite frequency allocation is from 3.7 to $4.2\,$GHz,
out of the C-BASS band. However, in spite of being attenuated by
$70\,$dB by the C-BASS bandpass filters
(Section~\ref{sec:bandpass_filters}), the out-of-band satellite broadcasts
were sufficiently strong that we were required to cascade two bandpass filters (BPFs) in
each signal chain to remove them from our data.

There are several terrestrial fixed links in the vicinity of the Owens
Valley Radio Observatory that are bright enough to interfere with our
observations. Since the northern receiver does not have a spectrometer backend, 
we use cryogenic notch filters tailored to the measured RFI
spectrum to deal with in-band interference. This comes at the price of
reduced receiver bandwidth (see Table~\ref{tab:passband_information}),
and complicated phase structure in the band (see Section~\ref{sec:results:passband}).

\begin{figure}
 \centering
 \includegraphics[width=0.47\textwidth]{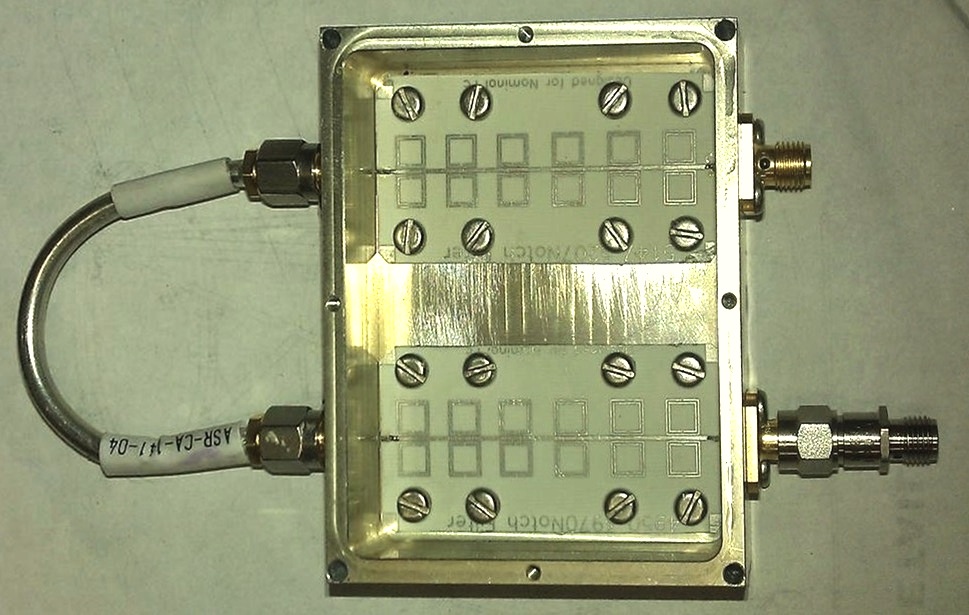}
 \caption{The C-BASS RFI notch filter assembly, consisting of two
   cascaded split-ring resonators housed in a single box. An isolator is placed
on the input to improve the matching.}
 \label{fig:notch_filter_photo}
\end{figure}

Ideally the notch filters should have a relatively high quality factor
$Q=\nu_{c}/\Delta \nu$, where $\nu_{c}$ is the centre frequency and $\Delta \nu$ the half-power
bandwidth of the notch. Each RFI band is approximately $25\,$MHz
wide (see Fig.~\ref{fig:rfi_environment}), which at $5\,$GHz
requires a $Q$-factor of 200. A quality factor this high requires
superconducting filters. However, since each of the
major RFI bands (i.e. $4.92\,$GHz and $5.18\,$GHz) seen at OVRO has a
second proximate RFI peak, a somewhat broader stop-band can be used. 
We designed two notch filters, each with a stop-band of
$80\,$MHz, to target these RFI bands. The notch filters are
split-ring resonators, which use the electromagnetic coupling between
resonant structures placed parallel to a microstrip transmission line
\citep{splitRingClassic,splitRingHTS}, and are fabricated using copper
on low-loss substrate \citep{CopleyPhD}. A picture of the filter assembly is shown in
Fig.~\ref{fig:notch_filter_photo}; both notch filters are housed in
a single box and an isolator is included on the input to improve the
matching.

The quality factor of the resonator is largely
determined by the resistivity of the conductor, so cooling the filters
results in a narrowing of the filter response and a
small shift in the resonant frequency, as shown in Fig.~\ref{fig:NotchFilterTempResponse}.  We
placed the filters on the $40\,$K stage of the cryostat, which gave a good match to the desired
stop-band.

\section{Polarimeter and radiometer} \label{sec:implementation_polarimeter}

The warm receiver components, along with the calibration noise diode,
are housed in a temperature controlled box beneath the cryostat and
above the dish surface. 
The gain and filter chains are shown as a simplified block diagram
in Fig.~\ref{fig:cbass_receiver_diagram}. They consist of isolators,
slope compensators, a bandpass filter, and multiple amplifiers. 
Below we describe the design of the
major receiver components; further details are in \citet{King:2009}.

\subsection{$180^{\circ}$ phase switches} \label{sec:phase_sw:hardware}

The phase switches were constructed using two broadband
surface-mount HMC547LP3 Single Pole Double Throw (SPDT) switches to
select between $0\dg$ and $180\dg$ signal paths with a rise time of 3\,ns.
The phase switch is powered by a $-5\,$V DC supply and can be switched by CMOS or TTL logic
signals. 

\begin{figure}
 \centering
\subfloat[]{
 \includegraphics[width=0.37\textwidth]{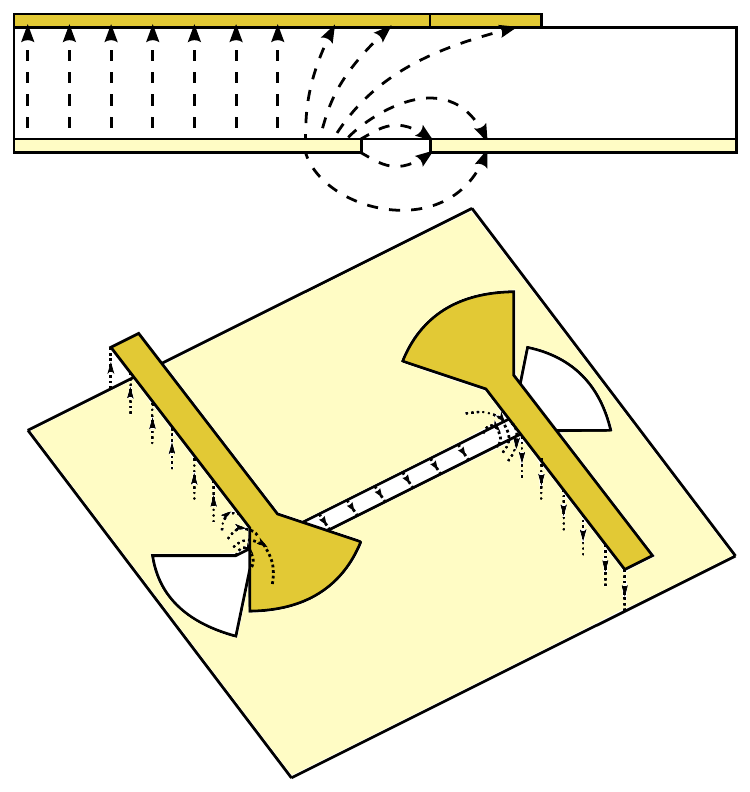}
}\\
\subfloat[]{
 \includegraphics[width=0.46\textwidth]{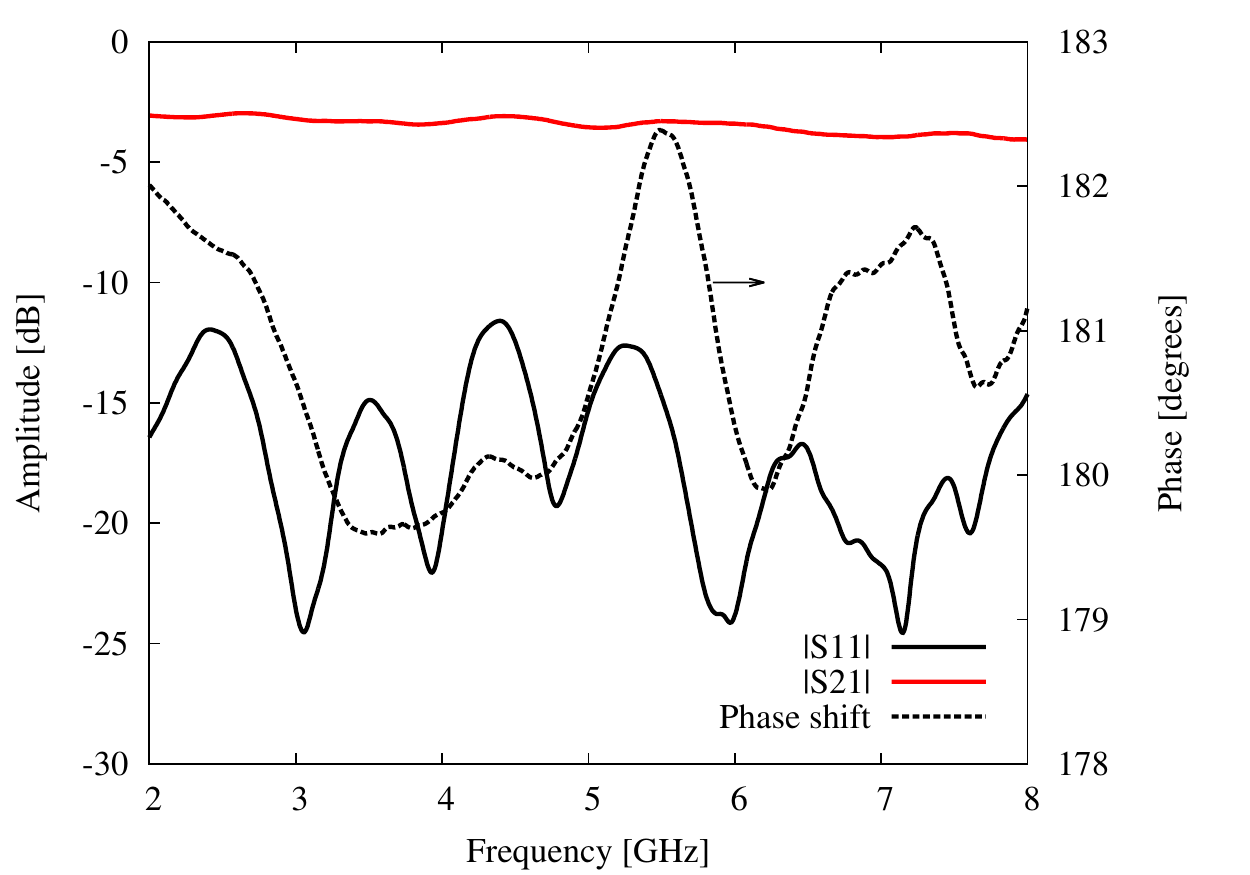}
}
 \caption{(a) Drawing of the transition used to achieve a $0\dg$ or
   $180\dg$ phase shift. The upper trace is coloured gold, the ground plane is coloured yellow,
and the electric field vectors are indicated by dashed lines. (b) The response of the phase
   switch. The induced phase shift is within a few degrees of $180\dg$
   from 2 to $8\,$GHz. The phase (indicated by the arrow) is plotted
   against the right-hand ordinate.}
 \label{fig:phase_switch_response}
\end{figure}

We obtain a $180\dg$ phase shift through geometric means by
using a microstrip to slotline transition. The electric field in the
microstrip line is coupled to a slotline in the ground plane. The
signal is then coupled back in to a microstrip line at the other end of
the slotline. Both microstrip lines and the slotline are terminated
with radial stubs. If the second microstrip line exits on the opposite
side of the slotline as compared to the input microstrip line then the
electric field is effectively flipped, and a $180\dg$ phase shift is
achieved. A drawing of the microstrip to slotline structure, and the
measured response of the phase switch, is shown in
Fig.~\ref{fig:phase_switch_response}.

\subsection{Warm amplifiers}

The warm amplifiers, which provide all of the gain after the LNAs,
are built around the Hittite HMC462LP5 broadband distributed
amplifier monolithic microwave integrated circuit (MMIC). Each amplifier block uses two stages of
this
self-biased, surface-mount low noise amplifier -- based on GaAs pseudomorphic
high-electron-mobility transistor (pHEMT)
technology -- to provide $\approx30\,$dB of gain from 2 to $20\,$GHz with a
noise temperature of $\approx200\,$K. The MMIC packages are
interconnected using grounded coplanar waveguide rather than microstrip, as this allows for better
broadband match between the devices and the SMA connectors, and a better grounding environment
for the MMIC packages.

\subsection{Bandpass filters} \label{sec:bandpass_filters}

The bandpass filters consist of two cascaded filters -- a
broadside edge-coupled bandpass filter (BPF) followed by a stepped
impedance lowpass filter (LPF) to remove the higher harmonic images of
the BPF response. The filters were fabricated on $0.254\,$mm thick RO4350
substrate ($\epsilon_r = 3.66$). The 3$\,$dB bandpass of the filter is
4.5 to $5.5\,$GHz. Each filter provides $\approx 70\,$dB of rejection in the
stop-band. This is not sufficient to reject the signals from the
brightest C-Band direct-broadcast satellites, so two complete filter
units are cascaded in each RF signal path.

\subsection{Detectors and digital backend} \label{sec:detectors_and_digital_backend}

The schematic diagram in Fig.~\ref{fig:cbass_receiver_diagram} shows the layout of the detection,
digitization, and digital signal processing scheme. We use zero-bias Schottky detector diodes, with
a video bandwidth of around $800\,$kHz, to measure the RF power. The detector diode voltages are
then filtered and digitized, as described in Section~\ref{sec:backend:filtering}. A digital signal
processing (DSP) chain (Section~\ref{sec:backend:dsp}) produces the final data stream from the raw
ADC values, which are transmitted over a USB link to the data acquisition computer.

The filter card and ADC/FPGA-based digital card are housed in the digital backend box
behind the telescope primary mirror. The cards are connected by a backplane,
powered by linear power supplies, and housed in shielded enclosures.

\subsubsection{Filtering and digitization} \label{sec:backend:filtering}

On the filter board the voltages are DC-coupled and low-pass filtered at $1\,$MHz, the
Nyquist frequency of the ADCs, by active voltage-control voltage-source (VCVS) anti-aliasing
filters. The filtered voltages are amplified by a factor of $\approx 6.5$ in order to
avoid digitization noise and maximize the dynamic range of the digitization. 

The digital card is designed around 16 ADCs capable of sampling at
$2.77\,$MHz and two Xilinx Spartan 3 FPGAs.  The card was originally designed for the Linear
Collider And Survey (LiCAS) particle physics experiment \citep{Reichold:2006wc}.  For the C-BASS
application, only 12 of the ADCs are used and sampling is performed at $2\,$MHz with 14$\,$bit
resolution. One of the FPGAs was re-programmed for C-BASS specific digital signal processing (DSP)
and control functions, while the other is an interface to on-board memory that is used for
diagnostic purposes.

\subsubsection{Digital processing} \label{sec:backend:dsp}

The DSP chain demodulates and integrates the twelve 2\,MHz data streams
to 10\,ms integrations. The raw data streams contain 60\,Hz mains pickup. 
In the DC-coupled mode used for diagnostic purposes that preserves the
full channel-by-channel information (see Section~\ref{sec:phase_sw}),
this signal and its harmonics are aliased to frequencies in the range
of the science data (0.01 -- 10\,Hz), which would put undesirable
contamination in the final images. 

For the science data stream, the
signals are first corrected for non-linearity of the detector diodes,
using look-up tables determined from laboratory measurements of the
individual diodes. The signals are then differenced to produce the
(sky $-$ load) signals, which also reduces the number of data streams
from 12 to 6. The 1\,kHz phase switch modulation is then demodulated,
which reduces the science signal to baseband while mixing the mains
and its harmonics to higher frequencies. A chain of decimating
low-pass fillters follows, reducing the data rate to 100\,Hz and
applying a 50\,Hz rectangular low-pass filter to the data. 

All these functions are carried out in the FPGA, which also produces the phase
switch drive signals and controls the noise diode. A USB interface
transmits the integrated data to the telescope control computer and
also allows control of the digital backend functions, such as firing
the noise diode.  A detailed description of the digital backend is
given in \citet{StevensonPhD}.

\section{Instrument performance} \label{sec:instrument_performance}

We have discussed the performance of individual components of the
receiver throughout the text. In this section we discuss the
performance of the receiver as a whole. This discussion is split in to
three sections: the variation of the receiver response with frequency
(i.e. the passband) in Section~\ref{sec:results:passband}, the stability of
the receiver in Section~\ref{sec:results:stability}, and the sensitivity of
the receiver in Section~\ref{sec:results:sensitivity}.

\subsection{Passband} \label{sec:results:passband}

The signal measured in each $I,Q,U$ polarization output is the integration of the instrument
response across the passband. 
In the DC-coupled diagnostic data stream we measure the antenna temperature and reference load temperature as two 
separate data streams, which we can write as (in this
case, for the left circular polarization (LCP) channel):
\begin{align}
r_{L} = & \int G_L(\nu) \left[ (1+\alpha_L(\nu) )T_{\rm L} +(1-\alpha_L(\nu) )T_{\rm ref} + T_N \right]d\nu \label{eqn:I_output}\\
r_{\rm ref} = & \int G_L(\nu) \left[ (1-\alpha_L(\nu) )T_{\rm L} +(1+\alpha_L(\nu) )T_{\rm ref} + 
T_N \right] d\nu. \label{eqn:ref_output}
\end{align}
Here $r_{L}$ and $r_{\rm ref}$ denote the recorded data values, $G_L(\nu)$ is the
frequency-dependent gain of the channel, and $\alpha_L(\nu)$ (which we call the imbalance parameter)
parametrizes the separation of the sky signal $T_L$ and the reference load signal $T_{\rm ref}$ by
the continuous
comparison radiometer architecture. $T_N$ is the noise temperature of
the receiver. The digital backend also records the filtered mode
differenced data stream $r_L - r_{\rm ref} \propto T_L
- T_{\rm ref}$. Ideally, $\alpha = 1$, i.e. the sky and load are
perfectly separated by the receiver.

\begin{figure}
 \centering
 \includegraphics[width=0.47\textwidth]{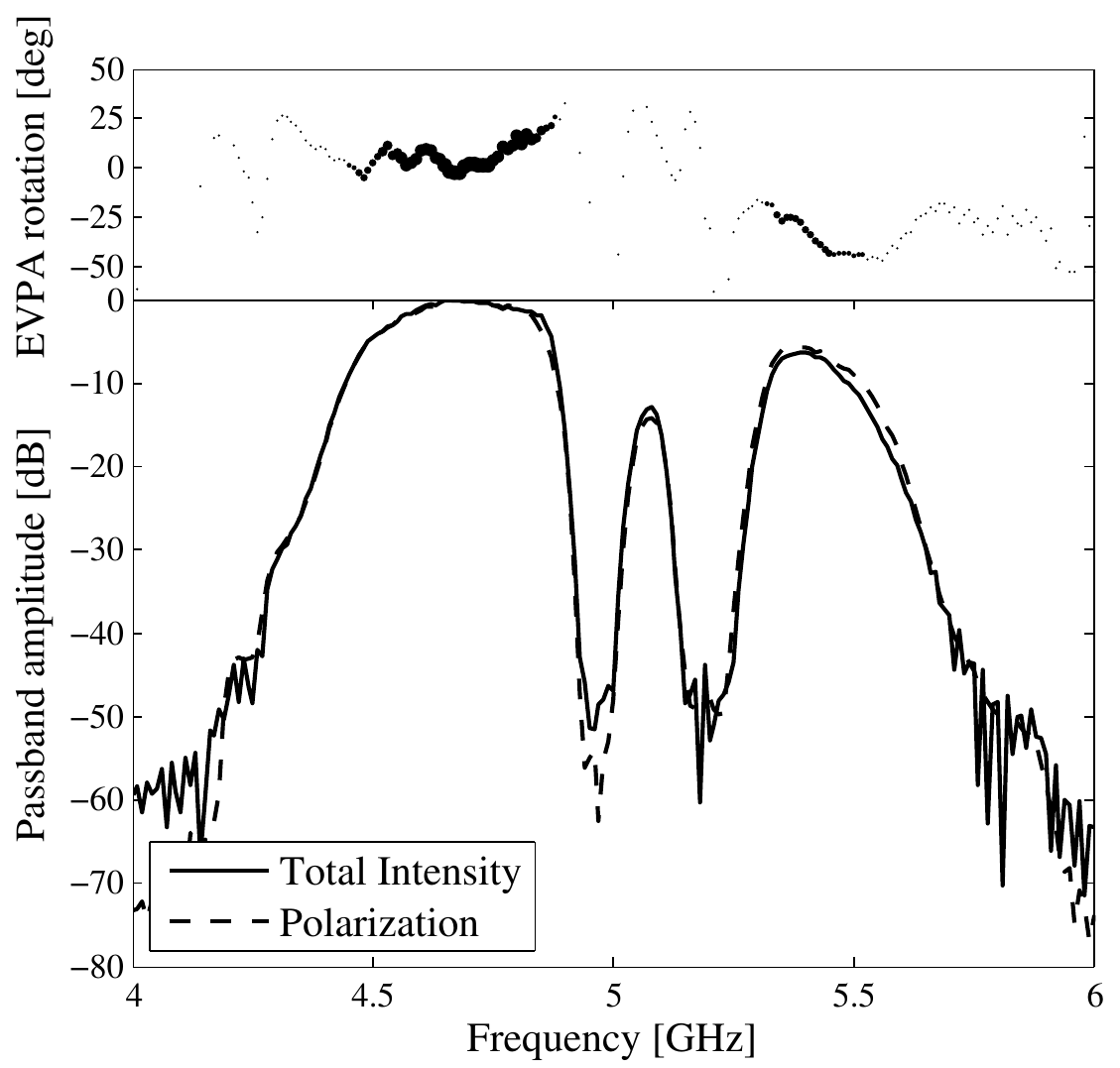}
\caption{(bottom) The passband amplitude of the total intensity (solid) and
  linear polarization (dashed) channels of the C-BASS instrument. The notches
  from the RFI filters are clearly seen in the centre of the band. 
(top) The EVPA rotation angle $\theta$ of the
instrument (Equations
  \ref{eqn:q_output} and \ref{eqn:u_output}). The size of each
  frequency point is proportional to the amplitude of the instrument
  response (lower plot) at that frequency.}
 \label{fig:cbass_passband}
\end{figure}

We can express the $Q$ and $U$ channels of the receiver in terms of
the input linear polarization vector $(Q_{\rm in},U_{\rm in})$,
in the reference frame of the OMT, as:
\begin{align}
r_Q = & \int G_P(\nu)\left[ \cos\theta(\nu)Q_{\rm in} - \sin\theta(\nu)U_{\rm in} \right] d\nu \label{eqn:q_output}\\
r_U = & \int G_P(\nu)\left[ \sin\theta(\nu)Q_{\rm in} + \cos\theta(\nu)U_{\rm in} \right] d\nu,
\label{eqn:u_output} 
\end{align}
where $G_P(\nu)$ and $\theta(\nu)$ are the frequency-dependent gain
and polarization angle (electric vector position angle, EVPA) rotation of the instrument
respectively. Ideally, for no loss in sensitivity, the instrument would
have a flat passband ($G_P(\nu) = {\rm constant}$ in the passband) and
a constant EVPA rotation angle ($\theta = {\rm constant}$).

We measure the passband of the northern C-BASS instrument by injecting
a sinusoidal voltage of known power with equal amplitude and phase
in to the noise diode injection ports. It appears as a purely linearly
polarized signal in the instrument reference frame. The frequency of
this signal is swept across the C-BASS band. At each frequency point
we turn the voltage signal generator on and off, and record the
response of the digital backend in each state. By taking the
difference of the receiver output in the on and off states we can
remove the baseline signal due to the system noise from
the data and directly measure the frequency-dependent response of the
receiver.

The effective centre frequency and noise-equivalent bandwidth of the
receiver are calculated according to the equations:
\begin{align}
\nonumber \nu_{c} = & \frac{\sum_i \nu_i G_i}{\sum_i G_i} \\
{\rm BW} = & \Delta \nu \frac{\left(\sum_i G_i \right)^2}{\sum_i G_i^2}.
\end{align}
Here $\nu_i,G_i$ are the frequency and gain of measurement $i$, and
$\Delta \nu$ is the width of each frequency bin (the gain is assumed
to be constant in each bin). This definition assumes a flat spectrum source
in power (or temperature). The parameters thus calculated are shown
in Table~\ref{tab:passband_information}. Both the total intensity and
linear polarization channels have very similar bandwidths and centre
frequencies. The bandwidth of the receiver has been reduced
by a factor of two by the notch filters.

\begin{table}
\centering
\caption{The centre frequency and noise-equivalent bandwidth of the
  total intensity and linear polarization channels, assuming a flat
spectrum source in power or temperature.}
\label{tab:passband_information}
\begin{tabular}{|c|c|c|}
\hline
 & Center freq. [GHz] & Bandwidth [GHz] \\ \hline
Total Intensity & 4.764 & 0.489 \\ 
Polarization & 4.783 & 0.499 \\ \hline
\end{tabular}
\end{table}

The passband of the instrument is shown in
Fig.~\ref{fig:cbass_passband}, where we plot the mean of the two total intensity channels and the
linear polarization intensity in the lower panel. In the upper panel we plot the EVPA rotation angle
$\theta$ from Equations \ref{eqn:q_output} and~\ref{eqn:u_output}. The
intensity of each point has been weighted by the amplitude
of the instrument response at that frequency.

We know from measuring the passband without the notch filters in place
that the large-scale variation in $\theta$ across the band -- the difference between the lower part
of the band and the upper part -- is due to the phase
structure in the notch filter transmission curves. This non-constant
value of $\theta$ across the band will reduce our sensitivity. 
We can estimate the effect of the variation in $\theta$ by calculating the ideal and actual
response of the instrument to a purely linearly polarized source. We
calculate an `ideal' instrument response by assuming $\theta(\nu)=0$
in Equations \ref{eqn:q_output} and \ref{eqn:u_output}. For a source
with a flat spectrum, i.e. $Q_{\rm in} = Q_0, U_{\rm in} = 0$ the
amplitude for the real instrument is 3.4\% lower than for the ideal
instrument. For a more realistic source spectrum of $Q_{\rm in} =
Q_0\nu^{-3}, U_{\rm in} = 0$ the real instrument response is 2.5\%
lower than the ideal instrument. This small reduction in polarization
sensitivity is an acceptable price to pay for removing the RFI
signature from the data.

\begin{figure}
\centering
\includegraphics[width=0.47\textwidth]{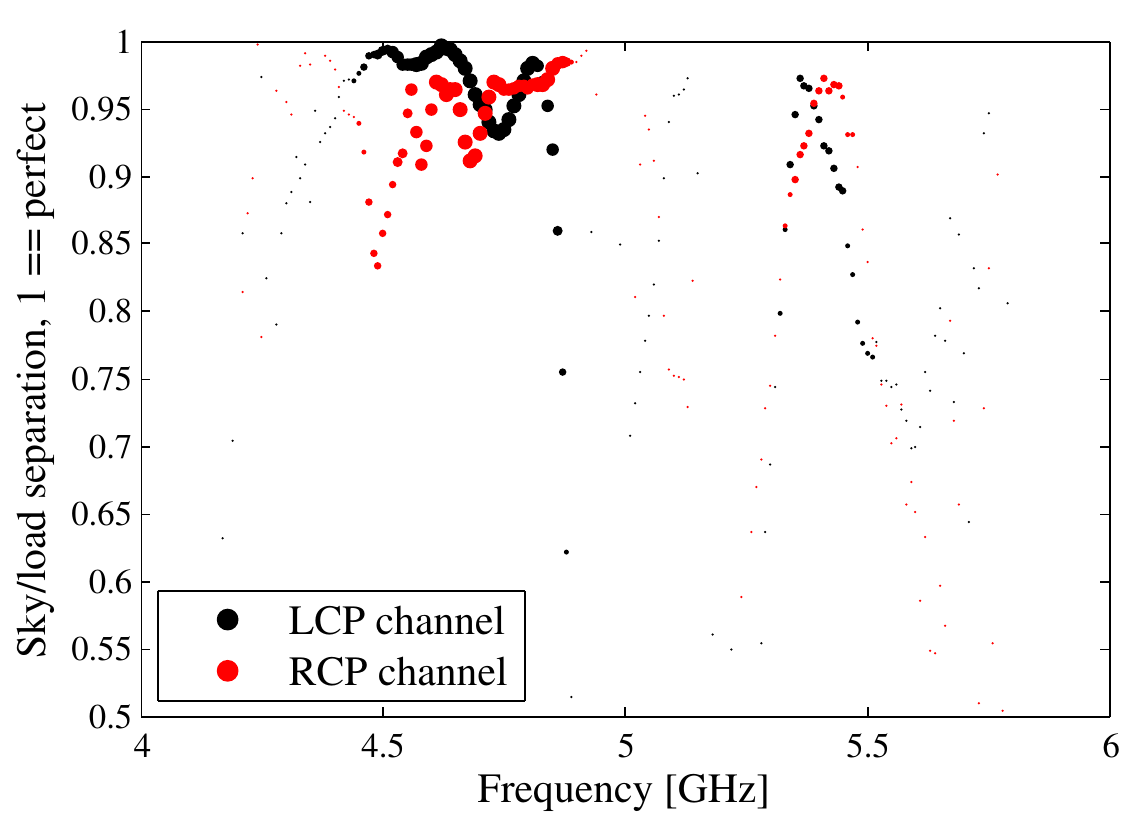}
\caption{The separation of sky and load by the continuous comparison
  radiometer in the total intensity LCP (black) and RCP (red) channels, as 
parameterized by $\alpha$ in Equations
  \ref{eqn:I_output} and \ref{eqn:ref_output}. A value of $\alpha = 1$
  indicates perfect separation, $\alpha = 0$ indicates no separation
  at all. The size of each frequency point is proportional to
  the amplitude of the instrument response at that frequency.}
\label{fig:cbass_passband_skyload}
\end{figure}

The sensitivity of the total intensity channel is affected by the
ability of the continuous comparison radiometer to properly separate
the sky and load signals at the input to the power detection stage in
Fig.~\ref{fig:cbass_receiver_diagram}. If $\alpha = 0$, i.e. the sky
and load signals are perfectly mixed, then $r_L-r_{\rm ref} = 0$
regardless of the sky temperature $T_L$: the signal is lost
entirely. The value of $\alpha$ for both continuous comparison
radiometers is shown in Fig.~\ref{fig:cbass_passband_skyload}; the
intensity of each frequency measurement has been weighted by the
instrument gain. The gain-weighted mean value of $\alpha$ is 0.96
for the LCP channel, and 0.95 for the RCP channel.

\subsection{Stability} \label{sec:results:stability}

\begin{figure}
 \centering
 \includegraphics[width=0.47\textwidth]{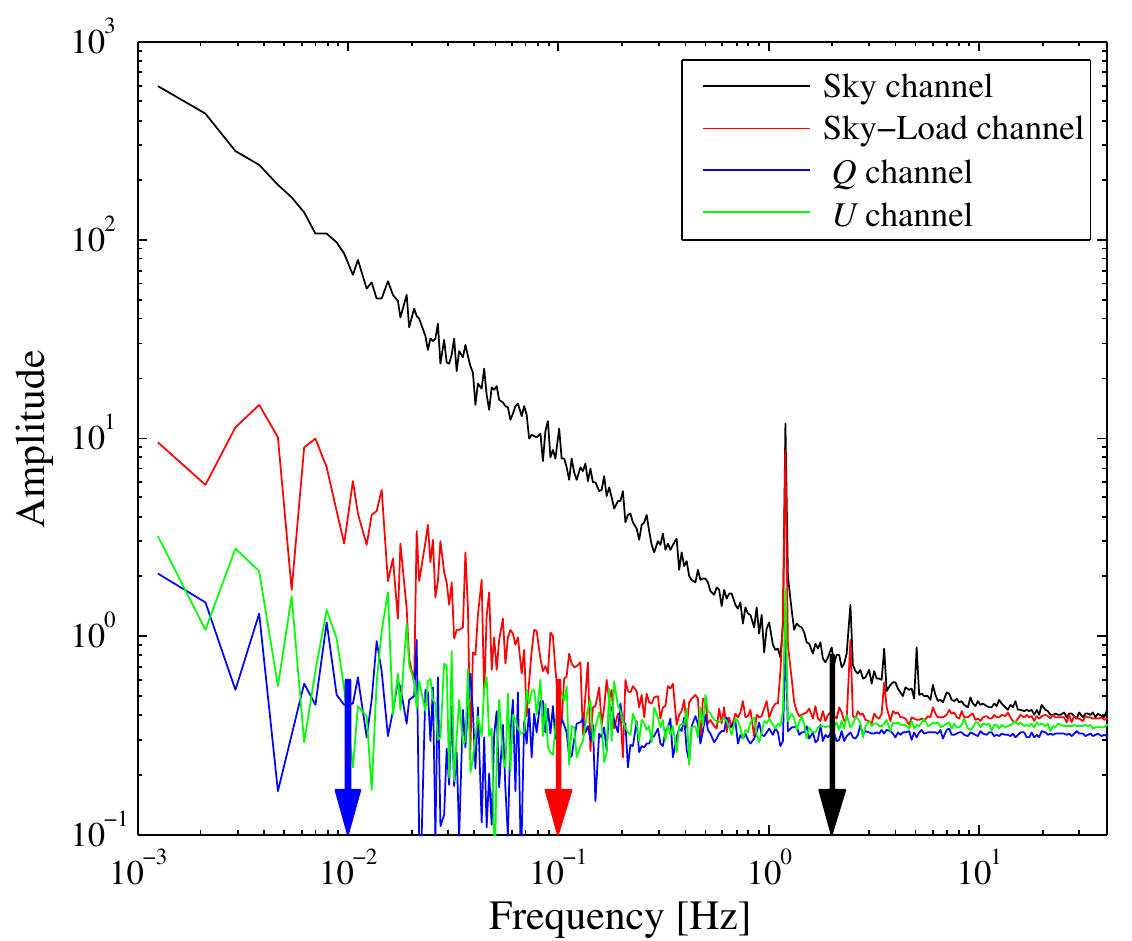}
\caption{Power spectra of data from the northern C-BASS receiver while
  observing the NCP for a continuous 20\,min time span. The data have
  had a mean subtracted but are otherwise uncalibrated. The strong
$1.2\,$Hz refrigeration 
  cycle signal and its harmonics are removed later by the data
  reduction pipeline. The black trace shows the sky data
  channel. The red trace shows the total intensity channel
  (sky$-$load). The knee frequencies are indicated by vertical arrows. The reduction of $1/f$ noise
by a factor of $\approx 20$ is testament to
  the power of the continuous-comparison radiometer architecture. The
  green and blue traces show the Stokes $Q$ and $U$ channels, with
  knee frequencies below $10\,$mHz. }
 \label{fig:power_spectrum}
\end{figure}

Fig.~\ref{fig:power_spectrum} shows the power spectrum of data taken
by the northern C-BASS instrument while observing the north celestial
pole (NCP) region where the astronomical signal remains constant due to our circularly
symmetric beam \citep{Holler:2012dp}. All the data plotted here are unprocessed and uncalibrated.
This is the only measurement presented in this paper based on astronomical data. 
The black trace shows the sky data channel from the DC-coupled data stream, which is analogous to
what a conventional radiometer would measure. The red trace shows the
total intensity ($T_L - T_{\rm ref}$) data channel from the filtered mode data stream. As expected, the level
of $1/f$ noise in the total intensity channel is reduced
compared to the sky channel. It is a factor of $\approx 20$ lower, and
the knee frequency has moved from $\approx 4\,$Hz to $\approx 0.2\,$Hz. The power spectrum
at $0.2\,$Hz is dominated by fluctuations in atmospheric emission, which we have measured to have a
typical knee frequency of $0.1\,$Hz.

The green and blue traces show the Stokes $Q$ and $U$ channels, both
with knee frequencies of below $10\,$mHz. This is sufficient given the typical 
scanning time of $\approx90\,$s ($11\,$mHz), and given that we remove large-scale fluctuations
using destriping map-making \citep{Sutton2009,Sutton2010}.
Fig.~\ref{fig:power_spectrum} demonstrates that both the
continuous-comparison radiometer and the pseudo-correlation
polarimeter sections of the C-BASS receiver remove the effect of receiver gain fluctuations on
the output signal.

\subsection{Sensitivity} \label{sec:results:sensitivity}

The receiver sensitivity is determined by the system temperature and
the passband. The radiometer design is a compromise between thermal
white noise level and low-frequency $1/f$ noise, as discussed in Section~\ref{sec:radiometer_design}; the
continuous-comparison design reduces the $1/f$ noise, but requires
there to be lossy components in front of the LNAs, which increase the
white noise level. We developed a receiver noise model based on the
Friis noise equation \citep{Friis:1944} to predict the noise temperature of the
receiver and verify its performance. The model uses measured or estimated losses and physical
temperatures for every component in the receiver, from the feed-horn
through to the second-stage (warm) amplifier. The noise contribution
of subsequent components is negligible. The model will be
described in detail in S.J.C. Muchovej et al. (in prep.).

The noise temperature was determined by measuring the ratio of powers
at the output of the cryostat when placing hot and cold loads in the
beam of the feed-horn, commonly known as a $Y$-factor $Y = P_{\rm
  hot}/P_{\rm cold}$ \citep{Kraus:1986}, where $P_{\rm hot}$ and $P_{\rm cold}$ are the measured
powers given a hot and cold load respectively. 
While the noise model is sufficiently complicated to warrant publication elsewhere, it is
constructive to consider a simplified case, as follows. For a simple receiver model in
which the total noise temperature can be modelled as the sum of a
receiver temperature $T_{\rm rx}$ and the hot or cold load
temperatures $T_{\rm hot}$ or $T_{\rm cold}$, the receiver
temperature can then be calculated from
\begin{align}
T_{\rm rx} = \frac{T_{\rm hot}-YT_{\rm cold}}{Y-1}.
\end{align}
For the continuous-comparison architecture the situation is
complicated by the fact that each RF output from the receiver is the
sum of voltages from a sky channel and the internal load. A simplified
model considers the input to the receiver as being the mean of the
load temperature and the temperature seen by the feed, resulting in
the modified relation
\begin{align}
T_{\rm rx} = \frac{ (T_{\rm hot}+T_{\rm load})/2 - Y(T_{\rm cold}+T_{\rm load})/2 }{Y-1}.
\label{eqn:y-factor}
\end{align}
This equation makes the assumption that the overall gain from the load
through to the receiver output is the same as that from the feed-horn aperture
to the output. This is not strictly true, and for detailed comparison
with the noise measurements we use the full receiver model. The
purpose of this equation is to present the reader with an idea of 
how the various parameters affect the receiver temperature.

We used the sky as the cold load with the telescope pointed at zenith, and 
a microwave absorbing foam at
ambient temperature as the hot load. The sky temperature at 5$\,$GHz is a
weak function of atmospheric conditions under typical conditions at
Owens Valley, and on a clear dry day is typically $\approx5\,$K (e.g. 
\citealt{1982RaSc...17.1455S}), including 2.7$\,$K
from the CMB. The precipitable water vapor at OVRO is typically 3 to 8\,mm on a clear day
\citep{PhilhourPhD}. We made two sets of measurements, one measuring the
receiver power using a power meter, with a bandpass filter to set the
same bandwidth as seen by the radiometer, the other using a spectrum
analyser to capture the power as a function of frequency. The former
method provides a close analogue of the power detected by the
broadband polarimeter/radiometer, while the latter allows us to check
for any anomalous behaviour as a function of frequency. 

\begin{figure}
\includegraphics[width=0.47\textwidth]{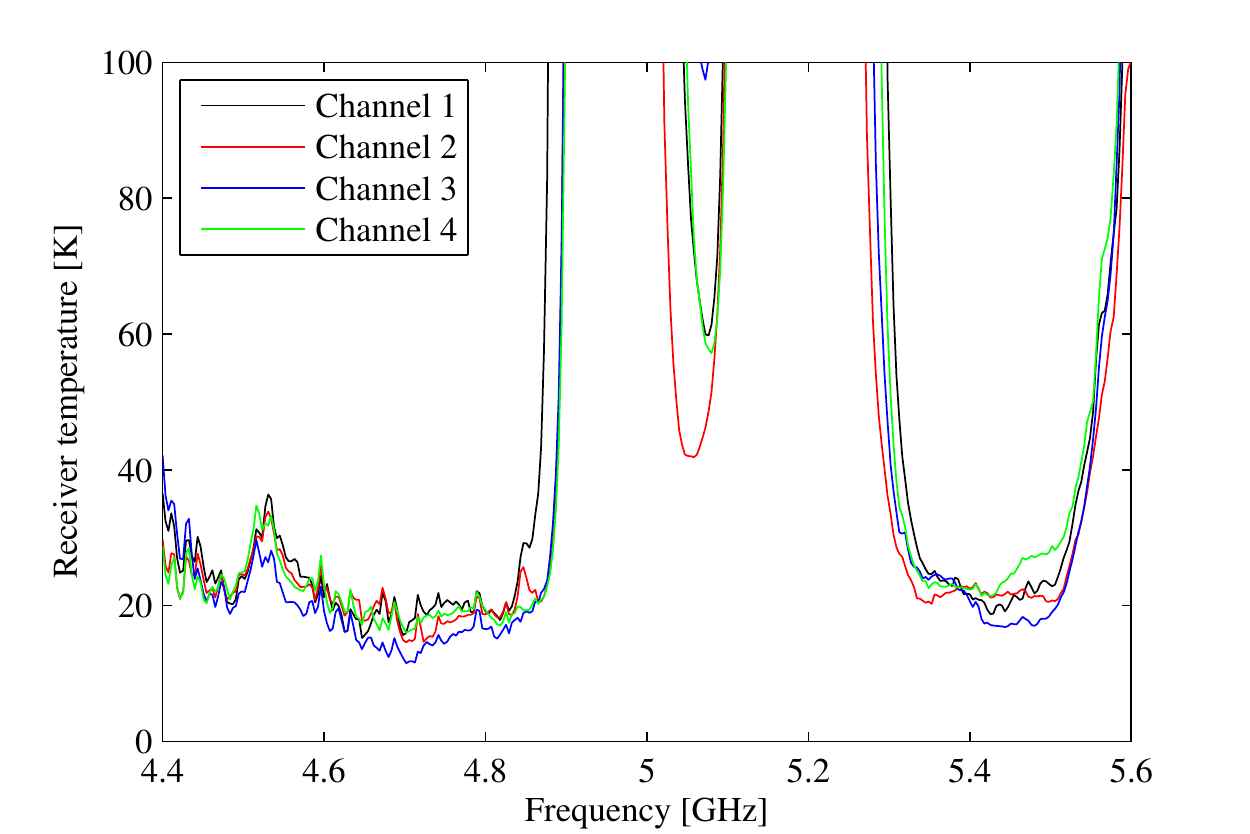}
\caption{Measured receiver temperature of the C-BASS receiver as a
  function of frequency, for each of the four cryostat RF output channels. The
  receiver temperatures are derived from power measurements taken with
  a spectrum analyser with hot and cold loads, using
  Equation~\ref{eqn:y-factor}. The effect of the notch filters can
  clearly be seen. The effective receiver temperature is the
  gain-weighted mean of this plot.  }
\label{figure:trx}
\end{figure}

The broadband $Y$-factor measurements for the four receiver output
channels without the secondary optics in place 
produce receiver temperatures, using Equation~\ref{eqn:y-factor} with $T_{\rm load} = 28\,$K, 
of 22, 20, 16 and 20$\,$K, with a mean value of
$19.5\pm2.2\,$K. The spectral measurements of the system temperature, calculated 
using Equation~\ref{eqn:y-factor}, are shown in Fig.~\ref{figure:trx}. 
A more accurate analysis using the full receiver model yields
a mean receiver temperature (not including the sky temperature and optics contribution) of
$16.0\pm1.8\,$K. 
This includes cold losses estimated from warm laboratory measurements of individual components, where we
find a loss for the sky channel before the hybrid of $-1.7\pm0.5\,$dB (comprising the OMT, the L2C,
the noise coupler, and
all their interconnecting cables), a loss between the cold load and
the hybrid of $-1.6\pm0.1\,$dB, and a cold loss common to both signal paths
(including the insertion loss of the hybrid itself, the isolators, and interconnecting cables) of
$-1.4\pm0.2\,$dB. The directly measured losses are quite uncertain (due to uncertainty in how they
change with temperature), so our best estimates of the losses come from the actual noise temperature
measurements.

The expected system temperature in polarization will be the receiver
temperature, plus the contribution from the sky (CMB and atmosphere),
the contribution from the telescope optics, and spillover. This is
expected to total $\approx 30\,{\rm K}$. 

The effective system temperature
in intensity is more complicated, since it has to take in to account
the noise contribution from the cold load as well as the sky
channel. Since each intensity output is the difference of the sky
channel and the load channel, and these have independent noise
signals, the noise level on one intensity output is $T_{\rm I} =
\left( T_{\rm sky\ channel}^2 + T_{\rm load\ channel}^2 \right)^{1/2}$, where
$T_{\rm sky\ channel} = T_{\rm sky} + T_{\rm spillover} + T_{\rm Rx\
  sky}$, and $T_{\rm load\ channel} = T_{\rm load} + T_{\rm Rx\
  load}$. $T_{\rm Rx\ sky}$ and $T_{\rm Rx\ load}$ are the receiver
contributions to the sky and load channels respectively; $T_{\rm Rx\
  sky} \neq T_{\rm Rx\ load}$ because of the different losses along
the two signal paths. The noise on the final summed intensity channel
is then reduced by a factor $\sqrt 2$ as the individual intensity
channels have independent noises. For a load temperature of $24\,{\rm
  K}$, which minimizes the $1/f$ fluctuations in the intensity data,
this gives an expected effective system temperature in intensity of
$32\,{\rm K}$.

\section{Conclusions}

The C-BASS project aims to produce all-sky maps of total intensity and linear polarization at
moderate resolution ($0.\!\dg73$) with a sensitivity of $0.1\,$mK per pixel. These will
be used primarily for removal of foregrounds from measurements of the
polarized CMB and will also be used, among other things, to study the Galactic magnetic field, to
constrain the frequency spectrum of the so-called `anomalous' emission and to study the Galactic
cosmic ray population.

We have described the instrument for the northern C-BASS survey which is a novel hybrid of a
continuous-comparison radiometer and a correlation polarimeter. A 
cryogenically cooled front end allows
sensitive measurements to be made. Most of the receiver components
have been specifically designed to provide excellent performance in our band. 
A digital backend digitizes and processes the detected powers.

We limit the level of undesirable spurious signals, such as pickup from the $60\,$Hz supply
voltage, by using phase switches to remove any low-frequency
fluctuations in the detector and digitization hardware. We
have shown that the continuous-comparison architecture works as
designed by reducing the $1/f$ noise seen in the total
intensity channel, and that the polarization channels show the
expected very low $1/f$ noise with a knee frequency of $10\,$mHz. 
The combination of low $1/f$ noise and a system temperature of $\approx 30\,$K in both total
intensity and linear polarization will allow us to reach our target sensitivity of $0.1\,$mK per
pixel.

\section*{Acknowledgements}

The C-BASS project is a collaboration between Caltech/JPL in the US,
Oxford and Manchester Universities in the UK, and Rhodes University
and the Hartebeesthoek Radio Astronomy Observatory in South Africa. It
is funded by the NSF (AST-0607857, AST-1010024, and AST-1212217), the University of Oxford, the
Royal Society, and the other participating institutions. We would like to
thank Russ Keeney for technical help at OVRO. We thank the Xilinx University Programme for their donation of FPGAs to
this project. OGK acknowledges the support of a Dorothy Hodgkin Award
in funding his studies while a student at Oxford, and the support of a
W.M. Keck Institute for Space Studies Postdoctoral Fellowship at
Caltech. CD acknowledges support from an STFC Advanced Fellowship, an
EU Marie-Curie IRG grant, and an ERC Starting Grant
(no. 307209). ACT acknowledges support from a Royal Society Dorothy Hodgkin Fellowship. 
CC acknowledges the support of the Commonwealth Scholarship, Square
Kilometer Array South Africa and Hertford College.


\label{lastpage}

\end{document}